\documentclass{emulateapj}

\usepackage{graphicx}
\usepackage{subfigure}
\usepackage{lscape}
\usepackage{apjfonts}
\usepackage{natbib}
\slugcomment{ApJ in Press; Received 2012 September 22; accepted 2012 November 20}

\newcommand{\angstrom}{{\rm \AA}}

\newcommand{\hbeta}{H{$\beta$}}
\newcommand{\halpha}{H{$\alpha$}}

\newcommand{\MgIIb}{Mg{\sevenrm\,II}\,$\lambda$2800}

\newcommand{\OIII}{[O{\sevenrm\,III}]}

\newcommand{\OIIIb}{[O{\sevenrm\,III}]\,$\lambda$5007}
\newcommand{\OIIIc}{[O{\sevenrm\,III}]\,$\lambda\lambda$4959,5007}

\newcommand{\NIIb}{[N{\sevenrm\,II}]\,$\lambda$6584}

 \font\sevenrm=cmr7 scaled 1000

\def\chandra{{\it Chandra}}

\newcommand{\lum}{\rm erg~s$^{-1}$}

\begin{document}

\title{Chandra X-ray and Hubble Space Telescope Imaging of
Optically Selected kiloparsec-Scale Binary Active Galactic
Nuclei I. Nature of the Nuclear Ionizing
Sources\altaffilmark{1}}

\shorttitle{kpc-Binary AGN I. Nature of the Ionizing Sources}

\shortauthors{LIU ET AL.}
\author{Xin Liu\altaffilmark{2,4}, Francesca Civano\altaffilmark{2},
Yue Shen\altaffilmark{2}, Paul J. Green\altaffilmark{2}, Jenny
E. Greene\altaffilmark{3}, and Michael A.
Strauss\altaffilmark{3}}

\altaffiltext{1}{Based, in part, on observations made with the
NASA/ESA {\it Hubble Space Telescope}, obtained at the Space
Telescope Science Institute, which is operated by the
Association of Universities for Research in Astronomy, Inc.,
under NASA contract NAS 5-26555. These observations are
associated with program number GO 12363.}

\altaffiltext{2}{Harvard-Smithsonian Center for Astrophysics,
60 Garden Street, Cambridge, MA 02138, USA}

\altaffiltext{3}{Department of Astrophysical Sciences,
Princeton University, Peyton Hall, Ivy Lane, Princeton, NJ
08544, USA}

\altaffiltext{4}{Einstein Fellow; xinliu@cfa.harvard.edu}

\begin{abstract}
Kiloparsec-scale binary active galactic nuclei (AGNs) signal
active supermassive black hole (SMBH) pairs in merging
galaxies. Despite their significance, unambiguously confirmed
cases remain scarce and most have been discovered
serendipitously. In a previous systematic search, we optically
identified four kpc-scale binary AGNs from candidates selected
with double-peaked narrow emission lines at $z=0.1$--0.2. Here
we present \chandra\ and {\it Hubble Space Telescope} Wide Field
Camera 3 (WFC3) imaging of these four systems. We critically
examine and confirm the binary-AGN scenario for two of the four
targets, by combining high angular resolution X-ray imaging
spectroscopy with \chandra\ ACIS-S, better nuclear position
constraints from WFC3 F105W imaging, and direct starburst
estimates from WFC3 F336W imaging; for the other two targets,
the existing data are still consistent with the binary-AGN
scenario, but we cannot rule out the possibility of only one
AGN ionizing gas in both merging galaxies. We find tentative
evidence for a systematically smaller X-ray-to-\OIII\
luminosity ratio and/or higher Compton-thick fraction in
optically selected kpc-scale binary AGNs than in single AGNs,
possibly caused by a higher nuclear gas column due to mergers
and/or a viewing angle bias related to the double-peak narrow
line selection. While our result lends some further support to
the general approach of optically identifying kpc-scale binary
AGNs, it also highlights the challenge and ambiguity of X-ray
confirmation.
\end{abstract}


\section{Introduction}\label{sec:intro}

\subsection{Significance of Binary Supermassive Black Holes}

Most bulge-dominated galaxies harbor central supermassive black
holes \citep[SMBHs;][]{kormendy95}. As a result,
binary\footnote{Following the initial nomenclature of
\citet{komossa03} for NGC 6240, we use ``binary'' AGNs to
denote a pair of AGNs, also in line with the nomenclature
``binary quasars'' adopted in the literature. In this context,
``binary'' does not necessarily presume that the black holes
themselves are gravitationally bound to each other (e.g., in
the case of kpc-scale binary AGNs, the host galaxies dominate
the potential well).} SMBHs are expected to form in galaxy
mergers \citep{begelman80,milosavljevic01,yu02}; they are an
inevitable and important consequence of the hierarchical bulge
and SMBH formation process. Binary SMBHs are believed to have a
significant dynamical impact on the nuclear stellar structure
of massive elliptical galaxies
\citep[e.g.,][]{faber97,ravindranath02,graham04,merritt06,kormendy09}.
The final inspiral and coalescence of hardened SMBH binaries
are predicted to produce strong gravitational wave signals
\citep{thorne76}, the detection of which would offer a direct
test of general relativity on cosmological scales
\citep{thorne87}. The identification and characterization of
binary SMBHs at various merger phases are valuable both for
understanding galaxy/SMBH evolution and for probing fundamental
physics \citep[see a comprehensive review by][]{colpi09}.

\subsection{Binary AGNs: the Kiloparsec Scales}

The frequency and statistical properties of binary active
galactic nuclei (AGNs) may offer useful insights to the
hierarchical merger paradigm of galaxy evolution \citep{yu11}
and the role of mergers in AGN fueling\footnote{While the
small-scale ($\sim0.1$--1 Mpc) quasar-quasar two-point
correlation function suggests a clustering excess over the
large-scale ($>1$ Mpc) extrapolation
\citep{hennawi06,myer07,hennawi09,shen10c}, it is still unclear
whether this is due to tidally enhanced BH accretion
\citep[e.g.,][]{djorgovski91,kochanek99,mortlock99}, or is
rather due to the small-scale clustering of their host dark
matter halos \citep[e.g.,][]{hopkins08,green11,richardson12}.
The projected separations of most of the observed binary
quasars are on scales of tens of kpc and larger, which may
still be too large for galaxy-galaxy tidal interactions to be
effective (but see \citealt{green10} for a counter-example of a
21-kpc separation binary quasar observed to have tidal features
indicative of ongoing interaction).}
\citep{ellison11,liu11b,silverman11,vanwassenhove12}. In a
galaxy merger, if both black holes (BHs) are simultaneously accreting, they
can be detected through spatially resolved emission diagnostics
which signal the presence of two AGNs. Theory suggests that
merger-induced gas inflows become significant (therefore likely
triggering AGNs) at separations under about a kpc
\citep{hernquist89}. Unlike the bound binary phase, which is
still extremely challenging to image directly
\citep[e.g.,][]{burke11}, the ``pairing'' phase (where the
separation between the two BHs, $a$, is a few tens pc to a few
tens kpc) is the most accessible because the two BHs are still
resolvable at cosmological distances (typical separation
$\gtrsim 1''$). Of particular interest is the late-pairing
phase ($a<10$ kpc), which connects mergers in a cosmological
context to pairs of BHs in galaxies, and sets the stage for the
subsequent evolution of close binaries.

\subsection{Systematic Searches for kpc-scale Binary AGNs}

The past few years have seen significant increase in the
inventory of kpc-scale binary AGNs, both from serendipitous
discoveries and from systematic searches. While the existence
of kpc-scale binary AGNs has been confirmed in a few pioneering
early discoveries \citep{moran92,komossa03,hudson06,bianchi08}
and further verified by more recent studies in X-rays
(\citealt{brassington07,fabbiano11,koss11,mazzarella11}; see
also \citealt{comerford11a} for a candidate), radio
\citep{fu11b,tadhunter12}, and optical broad emission lines
\citep{junkkarinen01,shields12}, the frequency of occurrence
and statistical properties of these systems remain poorly
constrained.

Addressing the frequency and statistical properties of binary
AGNs requires systematic searches. A natural approach is to
select candidates in galaxy mergers with double nuclei and
follow up to identify binary AGNs using diagnostic observations
such as X-ray imaging
(\citealt{guainazzi05,piconcelli10,koss12,teng12}) and/or
spatially resolved optical spectroscopy (\citealt{barth08};
\citealt{comerford09}\footnote{But see
\citet{civano10,civano12} and \citet{blecha12b} for an
alternative explanation for this particular candidate.};
\citealt{green10,greene11,liu11a,shields12}). However, because
of the requirement that the two nuclei are resolved in
ground-based optical imaging, systems identified using this
approach are in general biased against the late-pairing phase.

To mitigate this bias, an alternative approach is to select
candidates by kinematic signatures in spatially integrated
spectra, in analogy to the case of spectroscopic binary stars.
In particular, one such signature is the few hundred km
s$^{-1}$ velocity splitting observed in AGN narrow emission
lines \citep[NELs; e.g.,][]{sargent72,heckman81} such as \OIIIc
, which is seen in $\sim1$\% of low-redshift AGNs
\citep{liu10,smith09,wang09,ge12} as well as in quasars
\citep{shen11}. The working hypothesis is that the velocity
splitting signals the projected relative orbital motion of two
narrow line regions (NLRs), each ionized by its own central AGN
\citep[e.g.,][]{zhou04,gerke07,comerford08,xu09,peng11,barrows12}.
By selection, only binaries with projected angular separations
smaller than the spectroscopic aperture size will be included.
In principle, binary AGNs with separations as small as a few
tens of pc (limited by the intrinsic size of NLRs, which are
$\sim50$--1000 pc, and scale approximately as $L^{0.5}$;
\citealt{schmitt03,bennert02}) may be identified, if the
associated double stellar nuclei are resolvable by followup
observations using higher resolution near infrared (NIR)
imaging with {\it Hubble Space Telescope} ({\it HST}) and/or
ground-based adaptive optics (AO). Therefore, this kinematics
approach should be well suited for identifying binary SMBHs in
the late-pairing phase.

A major obstacle in identifying binary AGNs using the NEL
splitting signature, however, is that such profiles can also
arise from NLR gas kinematics around single AGNs, such as
rotating disks or bi-conical outflows
\citep{axon98,veilleux01,crenshaw09,shen10b,rosario10,fischer11,smith11,smith12}.
Nevertheless, our followup observations \citep{liu10b,shen10b}
of a subset of a systematically selected sample of 167 AGNs
with double-peaked NELs (\citealt{liu10}; see also
\citealt{smith09,wang09}), as well as studies by other groups
\citep{mcgurk11,fu11}, have demonstrated the feasibility and
importance of combining higher resolution NIR imaging
\citep{fu10,rosario11} and spatially resolved optical
spectroscopy \citep{comerford11b} to discriminate kpc-scale
binary AGNs from single-AGN-NLR gas kinematics. Kpc-scale
binary AGNs tend to show two concentrated \OIII\ nuclei
spatially coincident with two stellar bulges in a merger, with
the dynamics dominated by the potential of the individual
stellar bulges. Objects with complex NLR kinematics, in
contrast, usually exhibit bi-cone/disk shaped diffuse \OIII\
with a smooth single-peaked stellar background, as well
illustrated by the local example Mrk 78
\citep{whittle04,fischer11}. Roughly 10\% of the objects we
have followed up are best explained by binary AGNs at
(projected) kpc-scale separations.

\begin{deluxetable*}{lccccccccc}
\tabletypesize{\scriptsize} \tablecolumns{10}
\tablewidth{\textwidth}
%
\tablecaption{Kpc-scale binary-AGN candidates imaged with
HST/WFC3 and {\it Chandra} X-ray Observatory/ACIS.
\label{table:obs}}
\tablehead{\colhead{} & \colhead{} & \colhead{} & \colhead{} &
\colhead{Redshift} & \colhead{HST} & \colhead{$\Delta
\theta_{{\rm Y}}$} & \colhead{$r_{p, {\rm Y}}$} &
\colhead{Chandra}  &
\colhead{$\Delta \theta_{{\rm X-ray}}$} \\
\colhead{Target Name} & \colhead{Plate} & \colhead{Fiber} &
\colhead{MJD} & \colhead{$z_c$} & \colhead{Obs. UT} &
\colhead{($''$)} & \colhead{(kpc)} & \colhead{Obs. UT} &
\colhead{($''$)} \\
\colhead{(1)} & \colhead{(2)} & \colhead{(3)} & \colhead{(4)} &
\colhead{(5)} & \colhead{(6)} & \colhead{(7)} & \colhead{(8)} &
\colhead{(9)} & \colhead{(10)}
}
\startdata
SDSS J110851.04+065901.4   & 1004 & 182 & 52723 & 0.1816 & 20110513 & 0.70 & 2.1 &
20110210
& 0.82 \\
SDSS J113126.08$-$020459.2 &  327 & 394 & 52294 & 0.1463 & 20110524 & 0.70 & 1.8 &
20110211
& \nodata \\
SDSS J114642.47+511029.6   &  881 & 241 & 52368 & 0.1300 & 20110620 & 2.71 & 6.3 &
20110423
& 3.03 \\
SDSS J133226.34+060627.4   & 1801 & 250 & 54156 & 0.2070 & 20110310 & 1.50 & 5.1 &
20111128
& \nodata \\
\enddata
\tablecomments{Col. (1): SDSS names with J2000 coordinates
given in the form of ``hhmmss.ss+ddmmss.s''. Cols. (2)--(4):
SDSS spectroscopic plate number, fiber ID, and Modified Julian
Date. Col. (5): systemic redshift measured from stellar
continuum absorption features in the SDSS fiber spectra. Cols.
(6) \& (9): dates of the HST and \chandra\ observations. Cols.
(7) \& (8): projected angular and physical separation between
the double nuclei measured from HST $Y$-band images. Col. (10):
angular separation between the double nuclei measured from ACIS
X-ray images. See Table \ref{table:astrometry} for measured
positions of the individual nuclei.}
\end{deluxetable*}

\subsection{This Work: High-resolution Optical and X-Ray Imaging of Optically Identified
kpc-scale Binary AGNs}

Here and in a companion paper \citep[][hereafter Paper
II]{liu12b}, we present F336W/$U$- and F105W/$Y$-band images
obtained using Wide Field Camera 3 (WFC3) on board the {\it
HST}, and 0.5--10 keV X-ray images taken with the {\it Chandra
X-ray Observatory} \citep{weisskopf96} Advanced CCD Imaging
Spectrometer \citep[ACIS;][]{garmire03} of the four optically
selected kpc-scale binary AGNs identified by \citet{liu10b}.
Although our ground-based NIR imaging and spatially resolved
optical spectroscopy strongly suggest that these galaxy mergers
host binary AGNs \citep{liu10b,shen10b}, the case is not
watertight. While seven of the eight NEL nuclei in the four
galaxies are optically classified as Type 2 Seyferts, one \OIII
-faint nucleus is optically classified as either a Type 2
Seyfert, a LINER, or a LINER-H II composite, the latter two
cases of which may be due to starburst and/or shock heating
rather than AGN excitation
\citep[e.g.,][]{lutz99,terashima00,eracleous02}. More
importantly, even in the ``Seyfert-Seyfert'' cases, there could
be only one AGN, which ionizes gas in both merging components
\citep[e.g.,][]{moran92}. Photo-ionization arguments based on
spatially resolved optical spectroscopy were unable to rule out
this possibility \citep{liu10b}, given that the nuclear
separation is not much larger than the sizes of individual
NLRs, and the large systematic uncertainties in the electron
density measurements from diagnostic emission line ratios.

Our present work is motivated to further clarify these
ambiguities concerning the nature of the ionizing sources in
optically selected kpc-scale binary AGNs. {\it HST}/WFC3's
$Y$-band images allow us to get better positional priors to
resolve the closely separated double nuclei in X-ray imaging;
the $U$-band images offer constraints on spatially resolved
star formation activity in the host galaxy (Section
\ref{subsec:hst}). Utilizing \chandra\ ACIS's superb spatial
resolution and imaging spectroscopy capability in the X-rays
(Section \ref{subsec:chandra}), we put more direct constraints
on the intrinsic X-ray luminosity for each individual nucleus
in the merging galaxies than those estimates inferred
empirically from \OIIIb\ emission-line luminosity (Section
\ref{subsec:xraylumi}). Combined with constraints on the
contribution from star formation to the observed X-ray
luminosities estimated from $U$-band imaging (Section
\ref{subsec:sf}), we critically examine the purported
binary-AGN nature of our targets (Section \ref{subsec:nature}).
Combining the new X-ray observations with our previous optical
spectroscopy, we tentatively characterize the fraction of
optical binary AGNs that are weak or Compton-thick X-ray
emitters, and compare with the fraction among single AGNs
(Section \ref{subsec:x2oratio}). Since our targets were
selected in a systematic search, our results have general
implications for the general approach of identifying kpc-scale
binaries in double-peaked AGNs (Section \ref{subsec:imp_iden}),
the attributes and limitations of optical identification
compared to X-ray searches (Section
\ref{subsec:compare_withxray}), and the frequency of kpc-scale
binary AGNs (Section \ref{subsec:frequency}).

Throughout this paper, we assume a concordance cosmology with
$\Omega_m = 0.3$, $\Omega_{\Lambda} = 0.7$, and $H_{0}=70$ km
s$^{-1}$ Mpc$^{-1}$, and use the AB magnitude system
\citep{oke74}.

\section{Target Selection and Properties}\label{sec:target}

In Table \ref{table:obs}, we list basic photometric and
spectroscopic properties of the four binary-AGN candidates.
These candidates were discovered from a subset of a parent
sample of 167 Type 2 AGNs with double-peaked \OIIIc\ emission
lines \citep{liu10}. The parent sample was identified from
14,756 Type 2 AGNs optically selected from the spectroscopic
catalog of the Sloan Digital Sky Survey \citep[SDSS;][]{york00}
Data Release Seven \citep[DR7;][]{SDSSDR7}. We focused on Type
2 (i.e., obscured) AGNs which allow us to study the host galaxy
properties without much contamination from the AGNs. The
optical emission line ratios \OIIIb /\hbeta\ and \NIIb
/\halpha\ are characteristic of Type 2 Seyferts according to
the \citet{kewley01} criterion based on the \citet{bpt}
diagnostics. We conducted ground-based deep NIR images and
optical slit spectra from the Magellan 6.5 m and the Apache
Point Observatory 3.5 m telescopes \citep{liu10b,shen10b}. We
identified four strong kpc-scale binary AGN candidates out of
43 objects observed\footnote{See also a fifth candidate, SDSS
J1356+1026, reported by \citet{shen10b} and
\citet{greene11,greene12}.} \citep{liu10b}. In each system, the
NIR images reveal tidal features and double stellar components
with a projected separation of several kpc, while optical slit
spectra show two Type 2 Seyfert nuclei (except for one nucleus
which could also be a LINER or a composite) spatially
coincident with the stellar components, with line-of-sight
velocity offsets of a few hundred km s$^{-1}$. In Table
\ref{table:astrometry} we list redshift and \OIIIb\ (hereafter
\OIII ) luminosity measurements for each individual nucleus
from our ground-based longslit spectroscopy.

\begin{figure*}
  \centering
    \includegraphics[width=55mm]{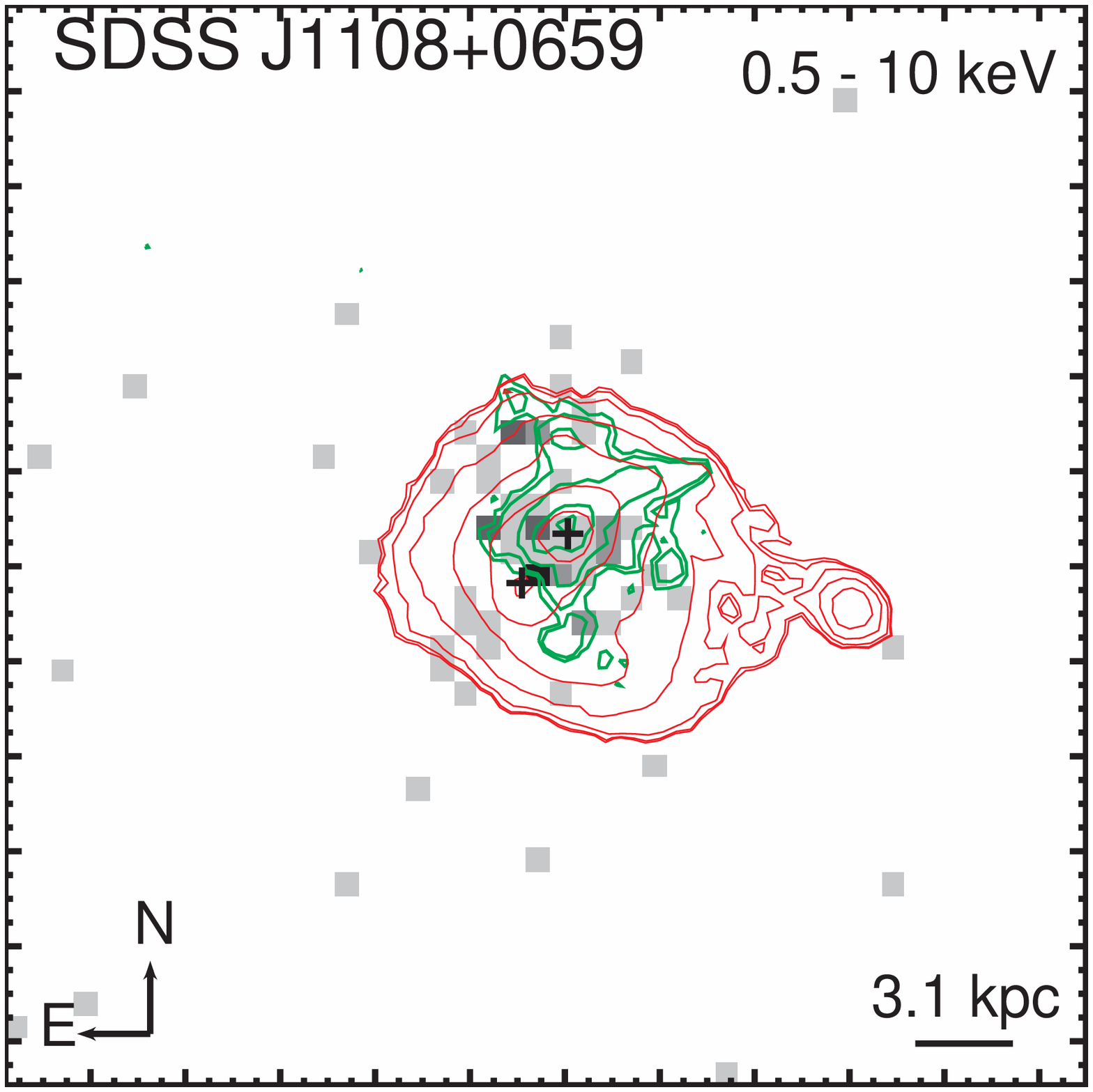}
    \includegraphics[width=55mm]{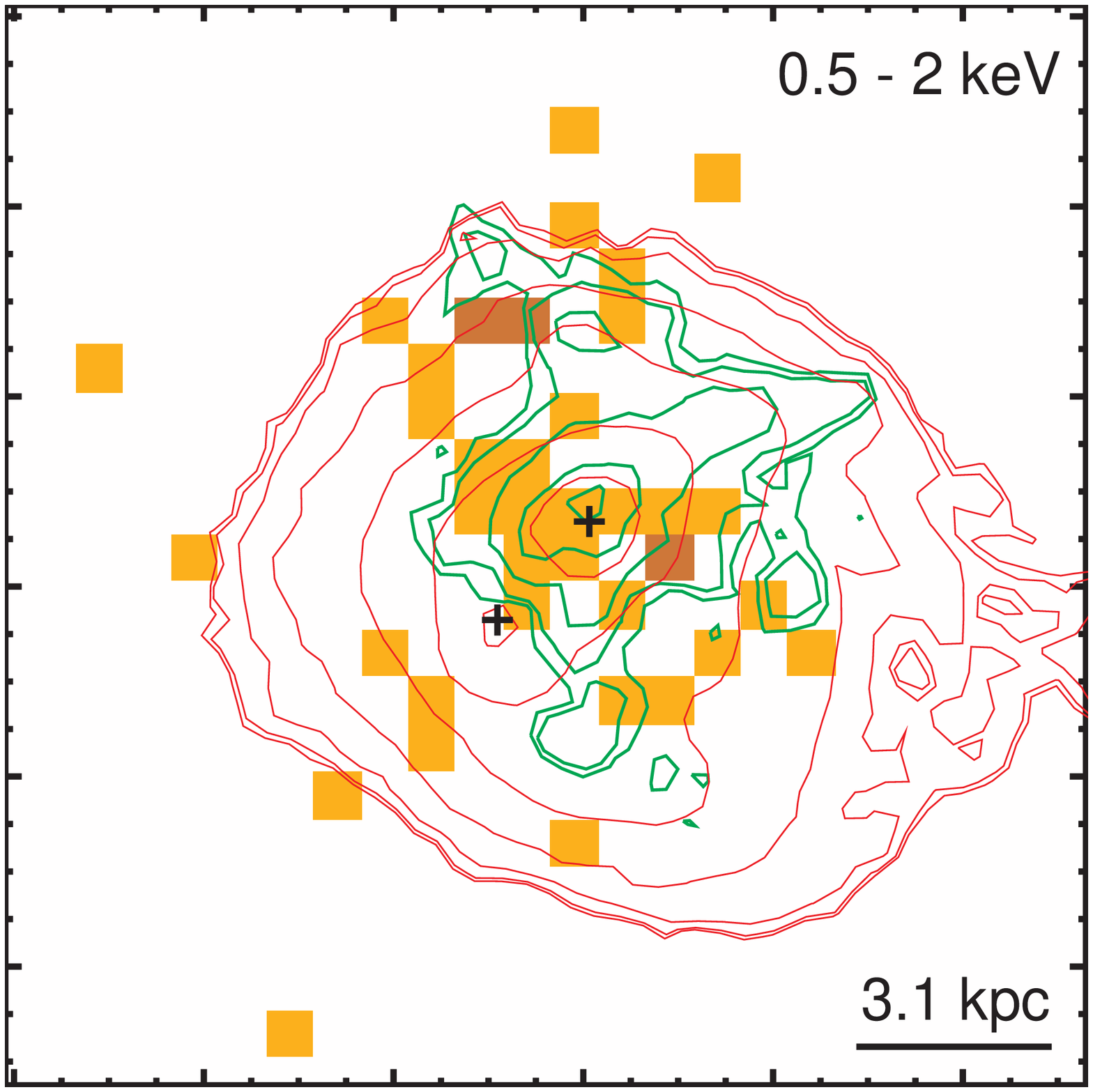}
    \includegraphics[width=55mm]{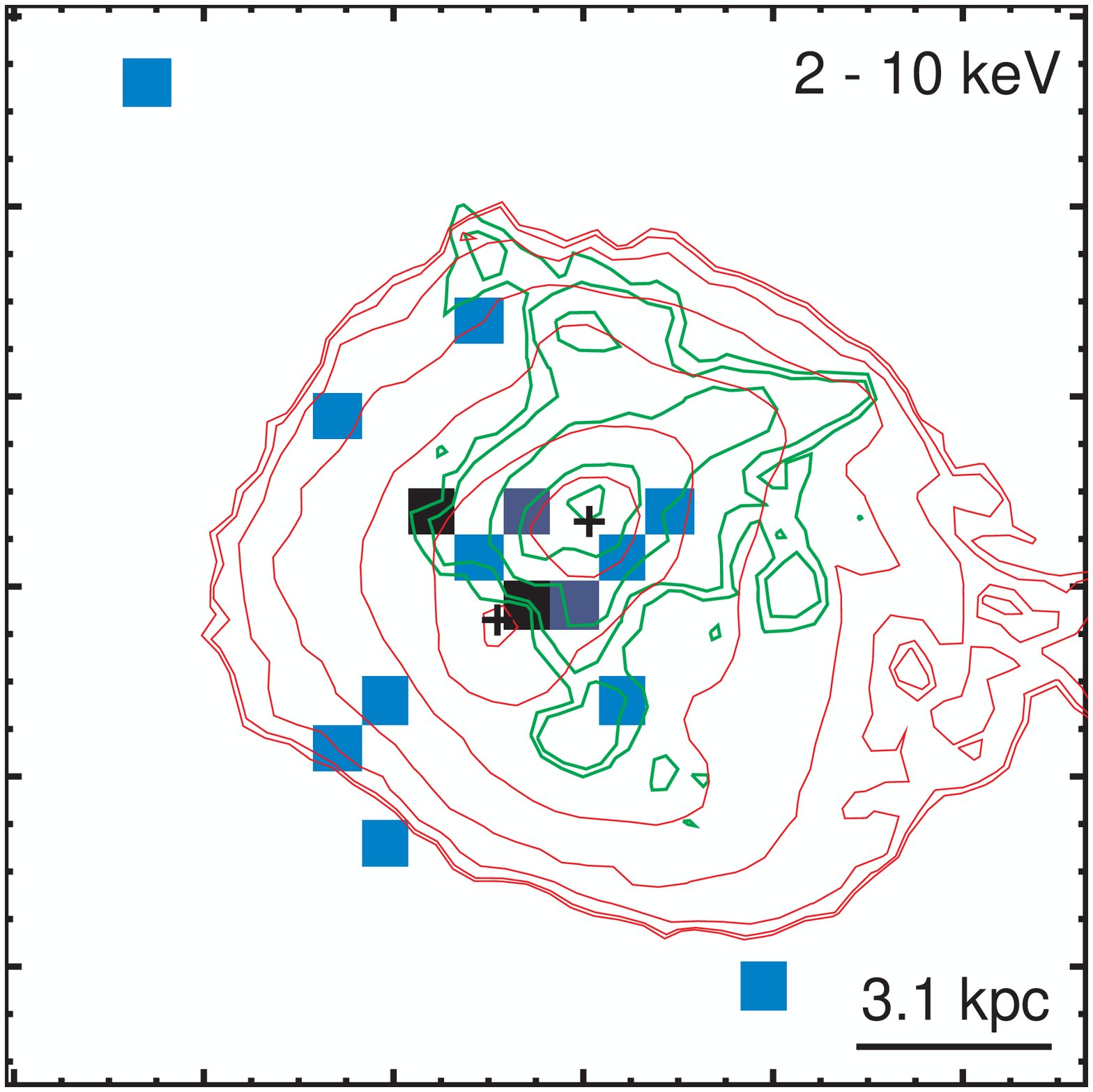}
    \includegraphics[width=55mm]{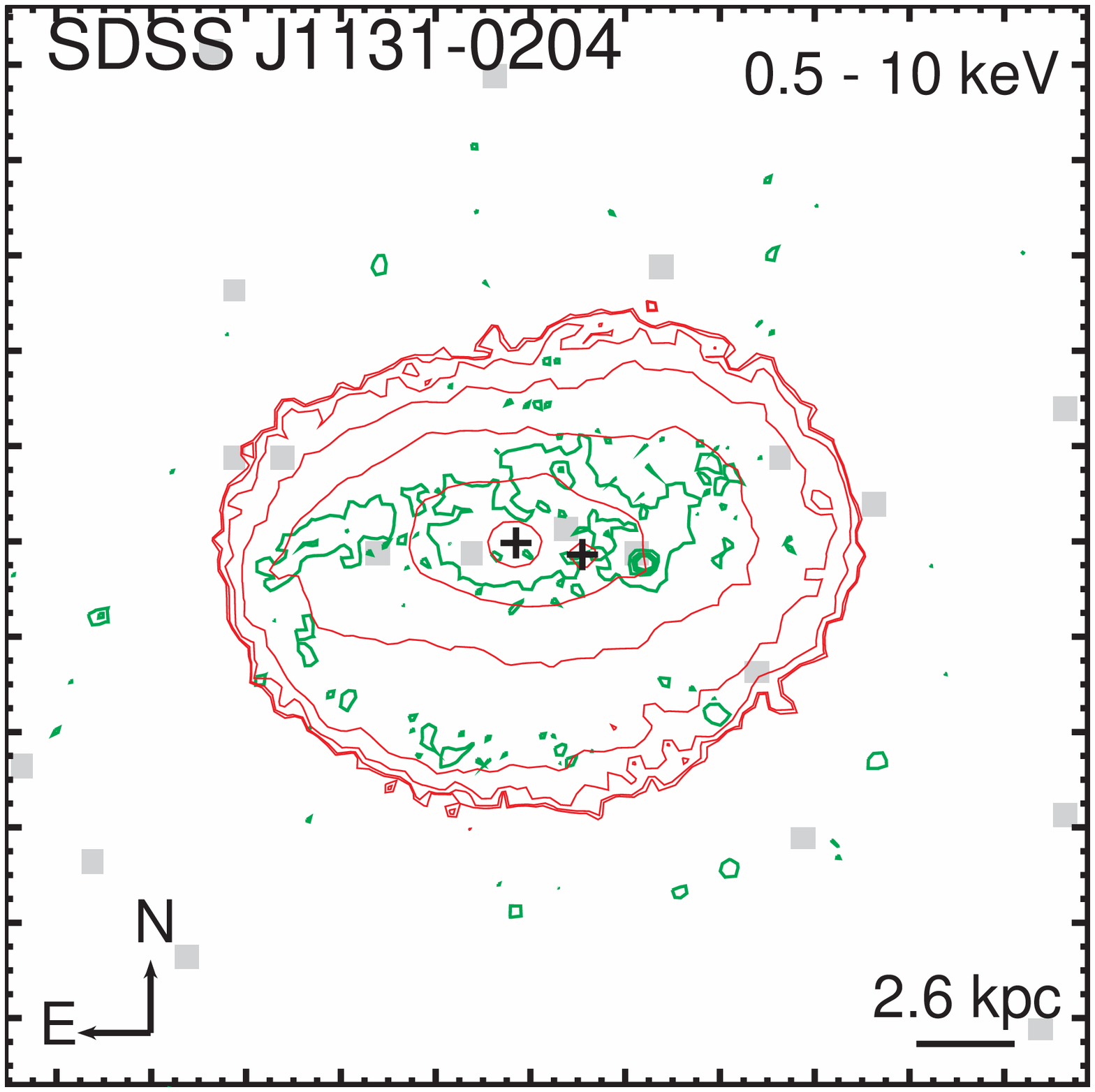}
    \includegraphics[width=55mm]{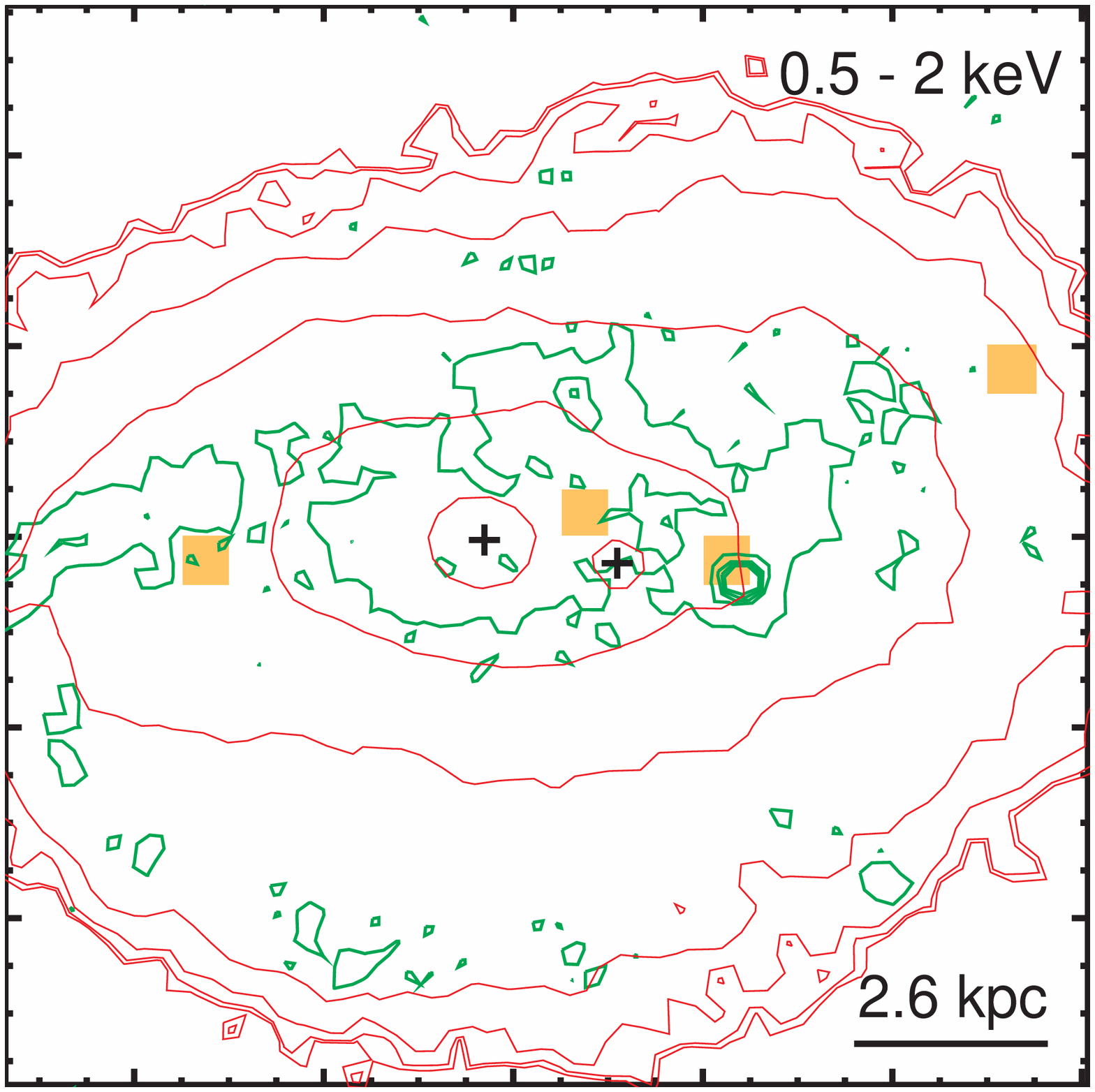}
    \includegraphics[width=55mm]{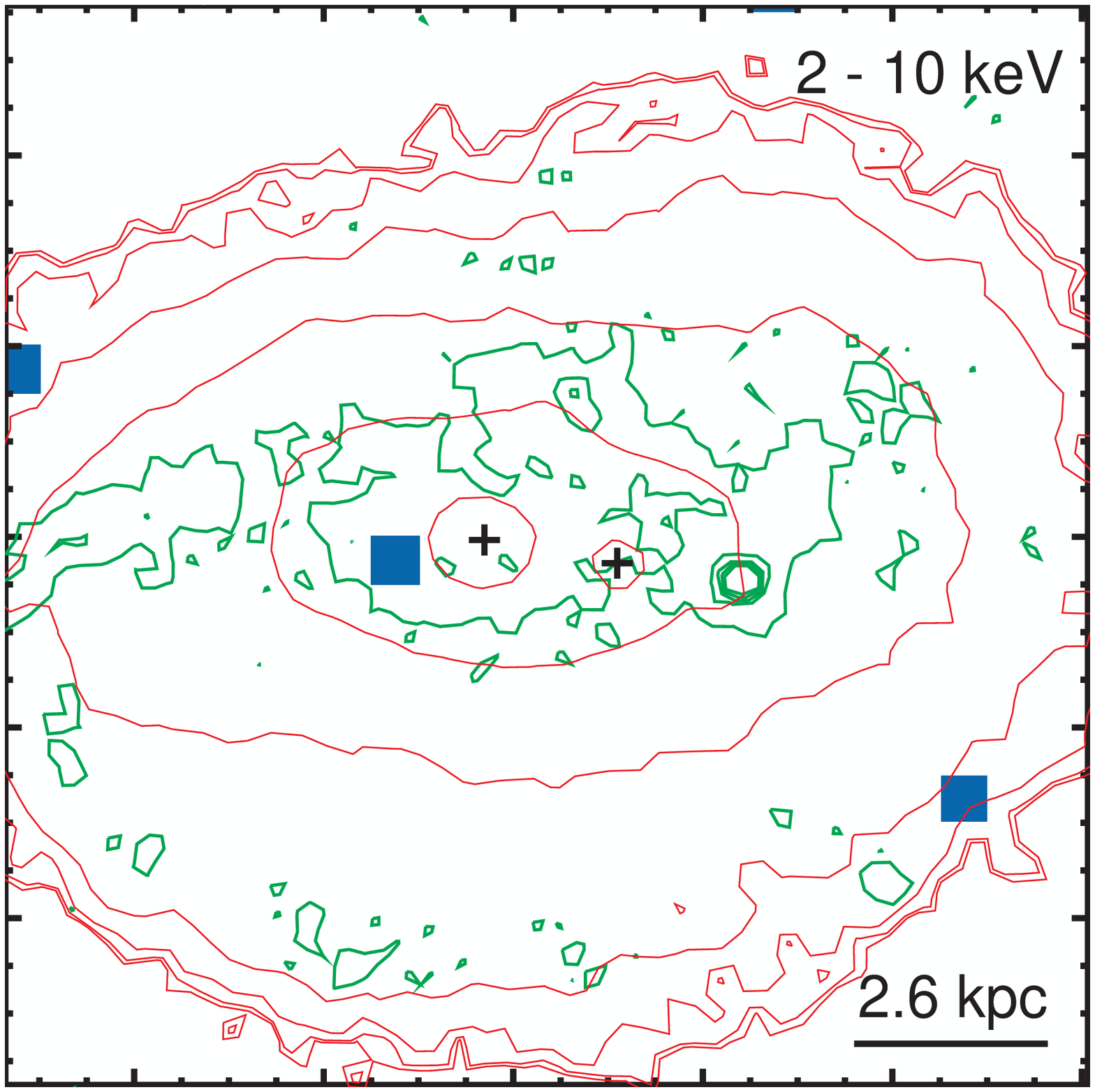}
    \includegraphics[width=55mm]{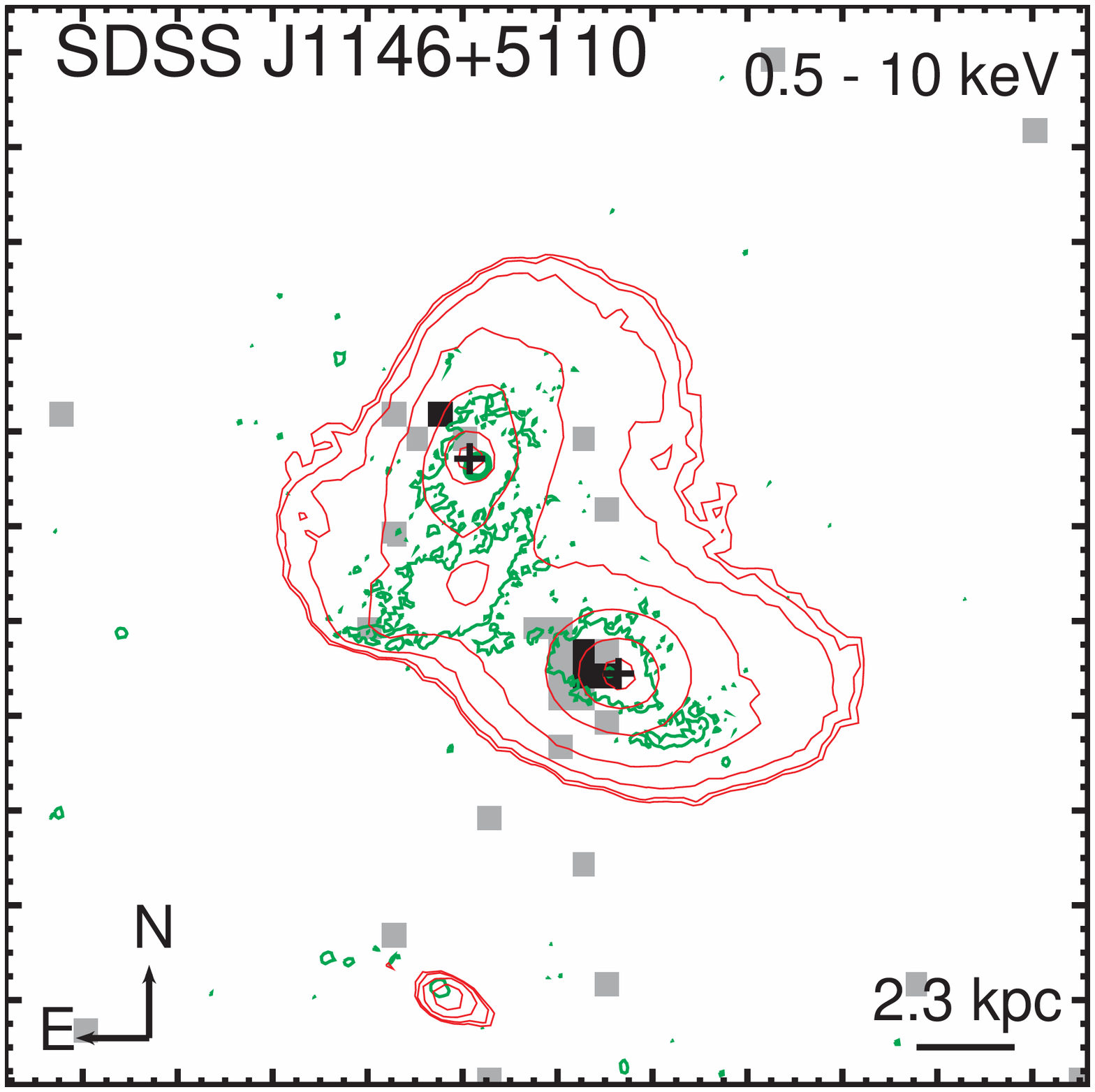}
    \includegraphics[width=55mm]{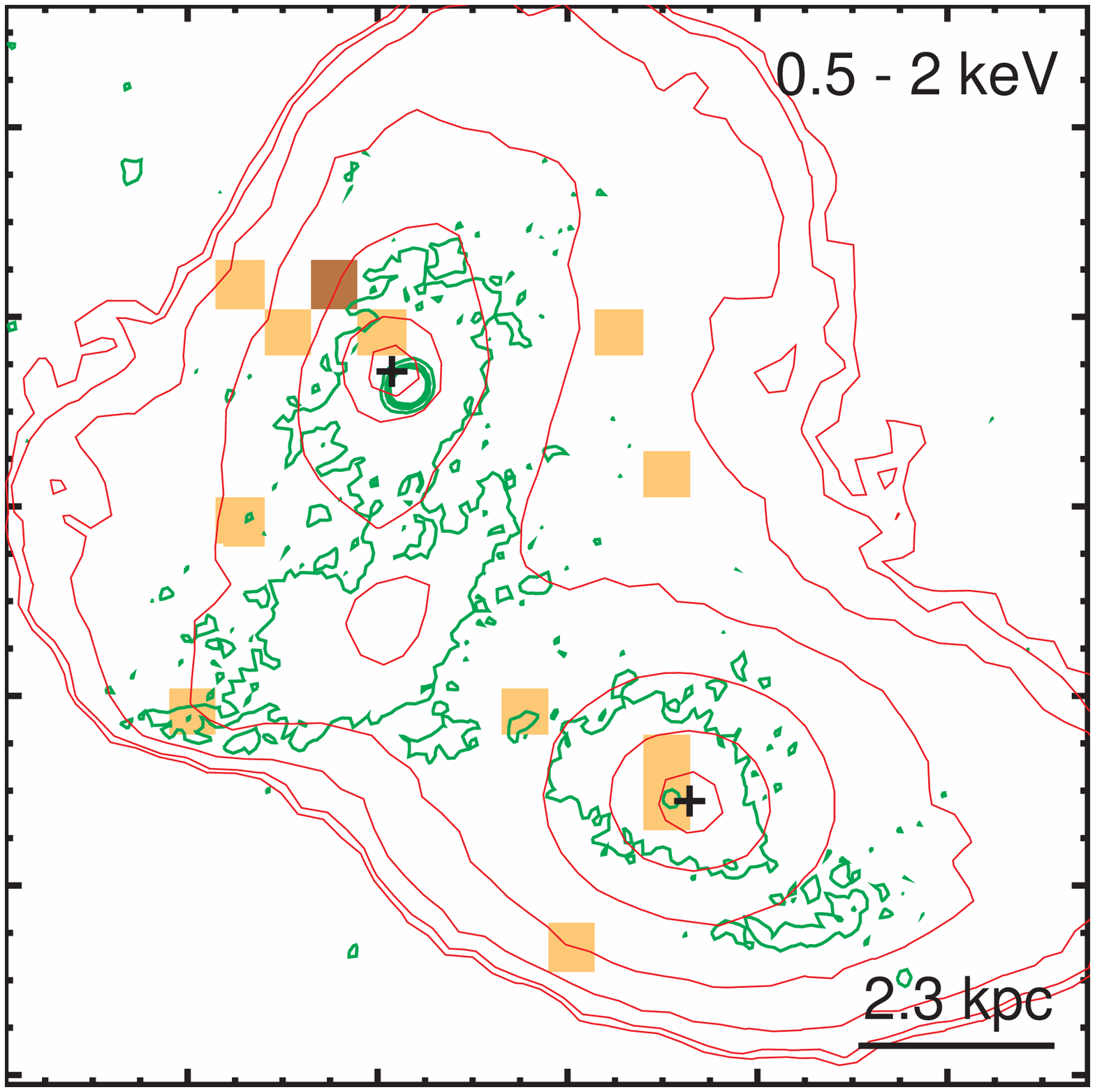}
    \includegraphics[width=55mm]{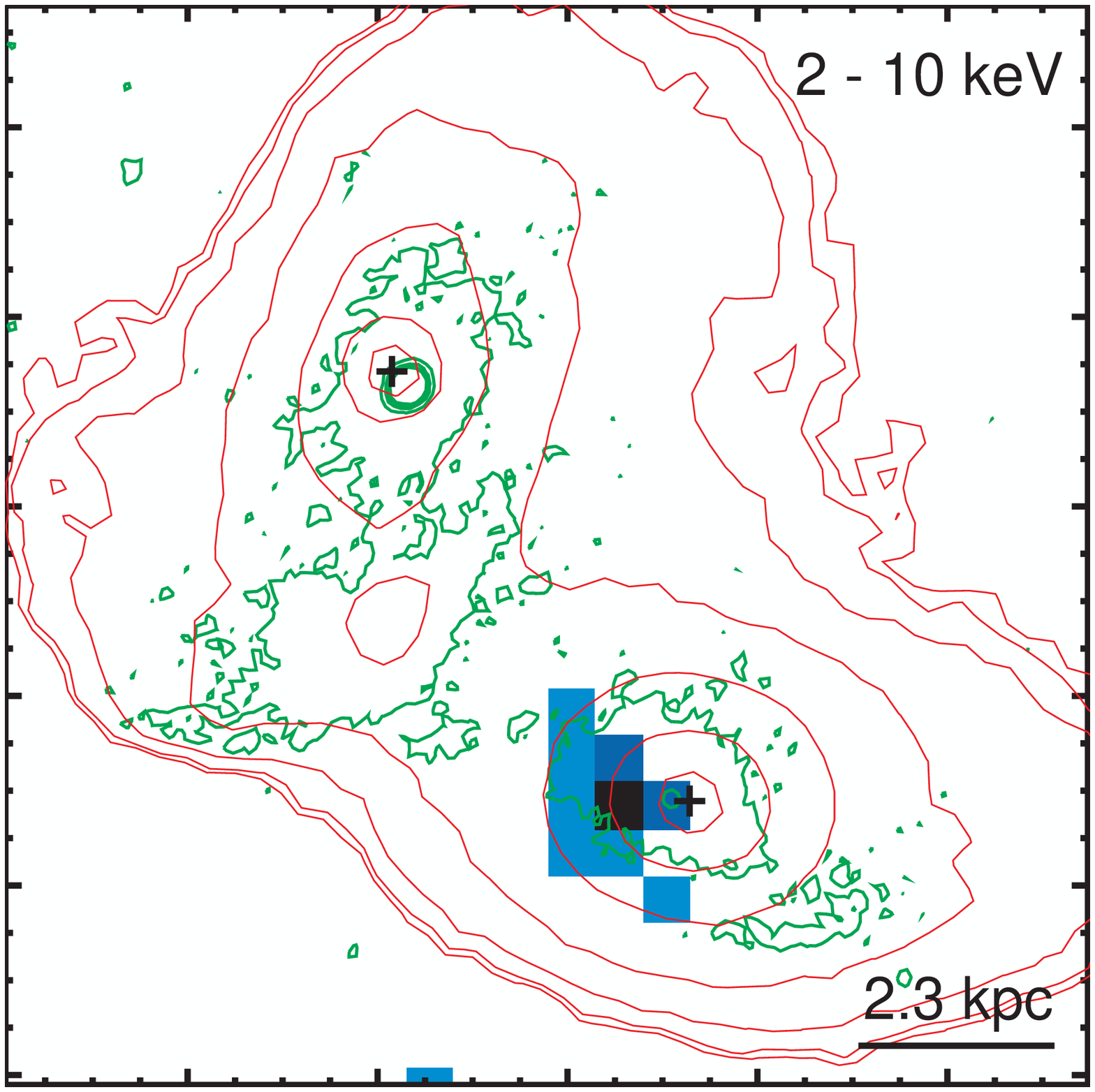}
    \includegraphics[width=55mm]{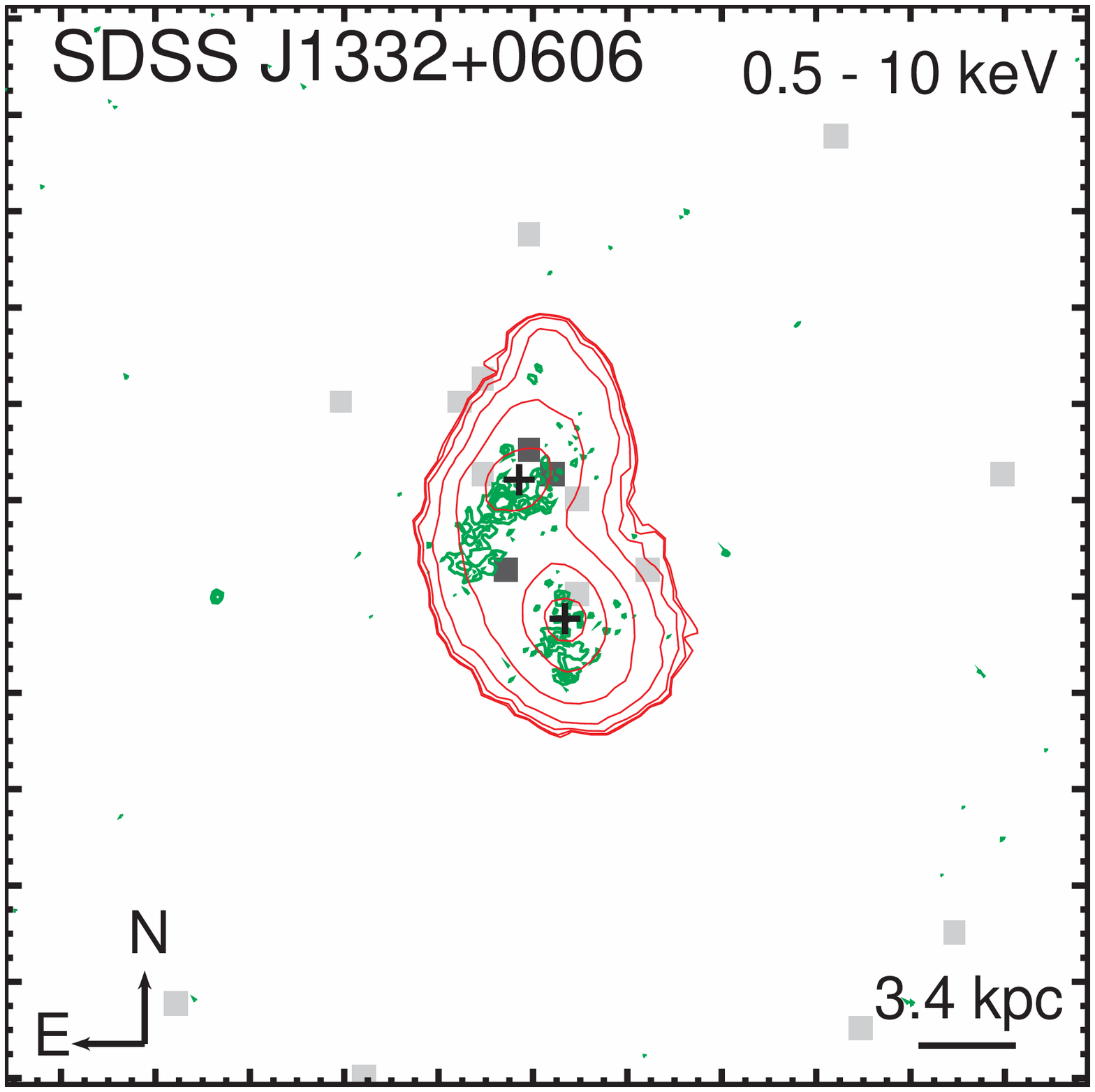}
    \includegraphics[width=55mm]{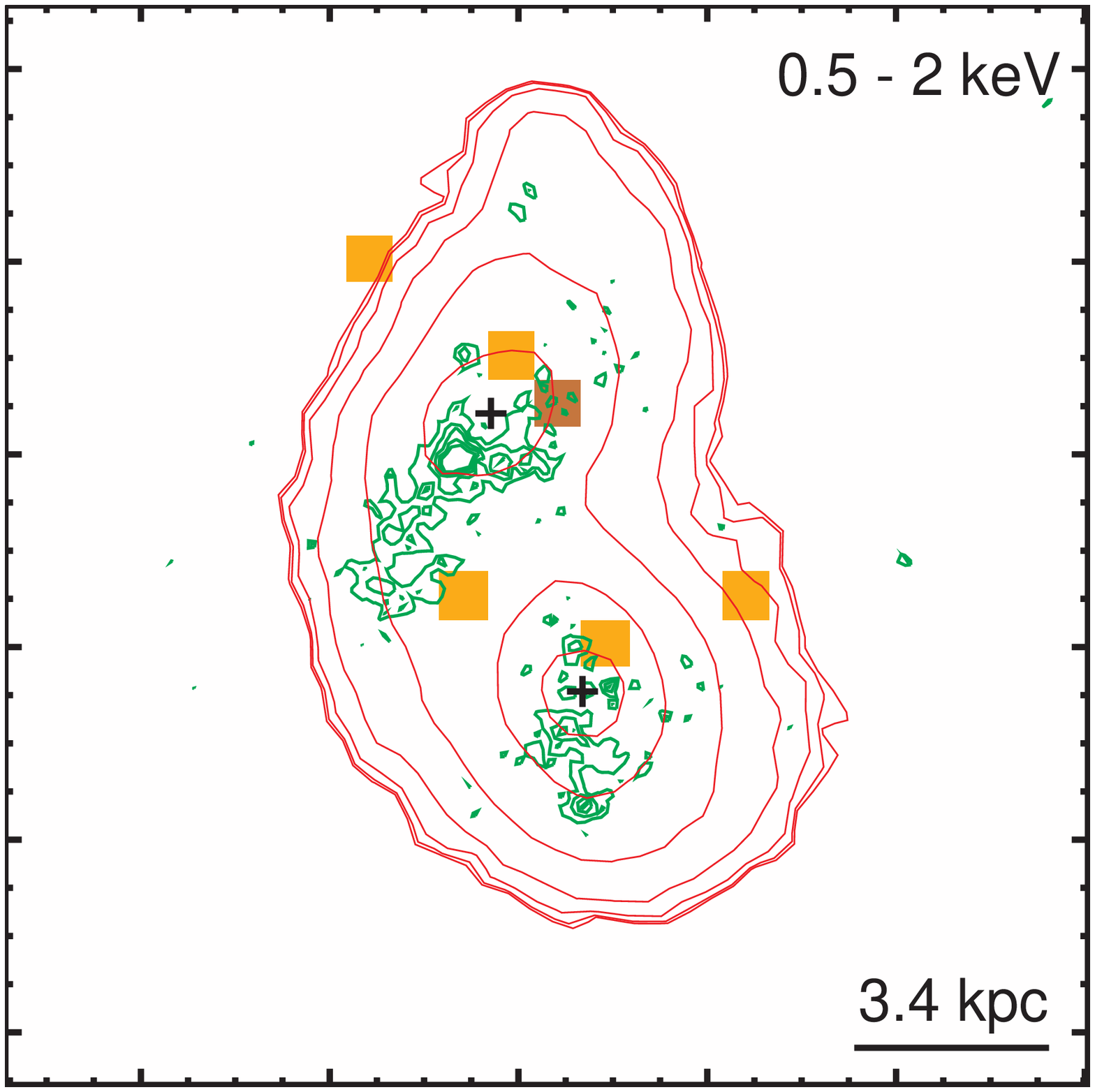}
    \includegraphics[width=55mm]{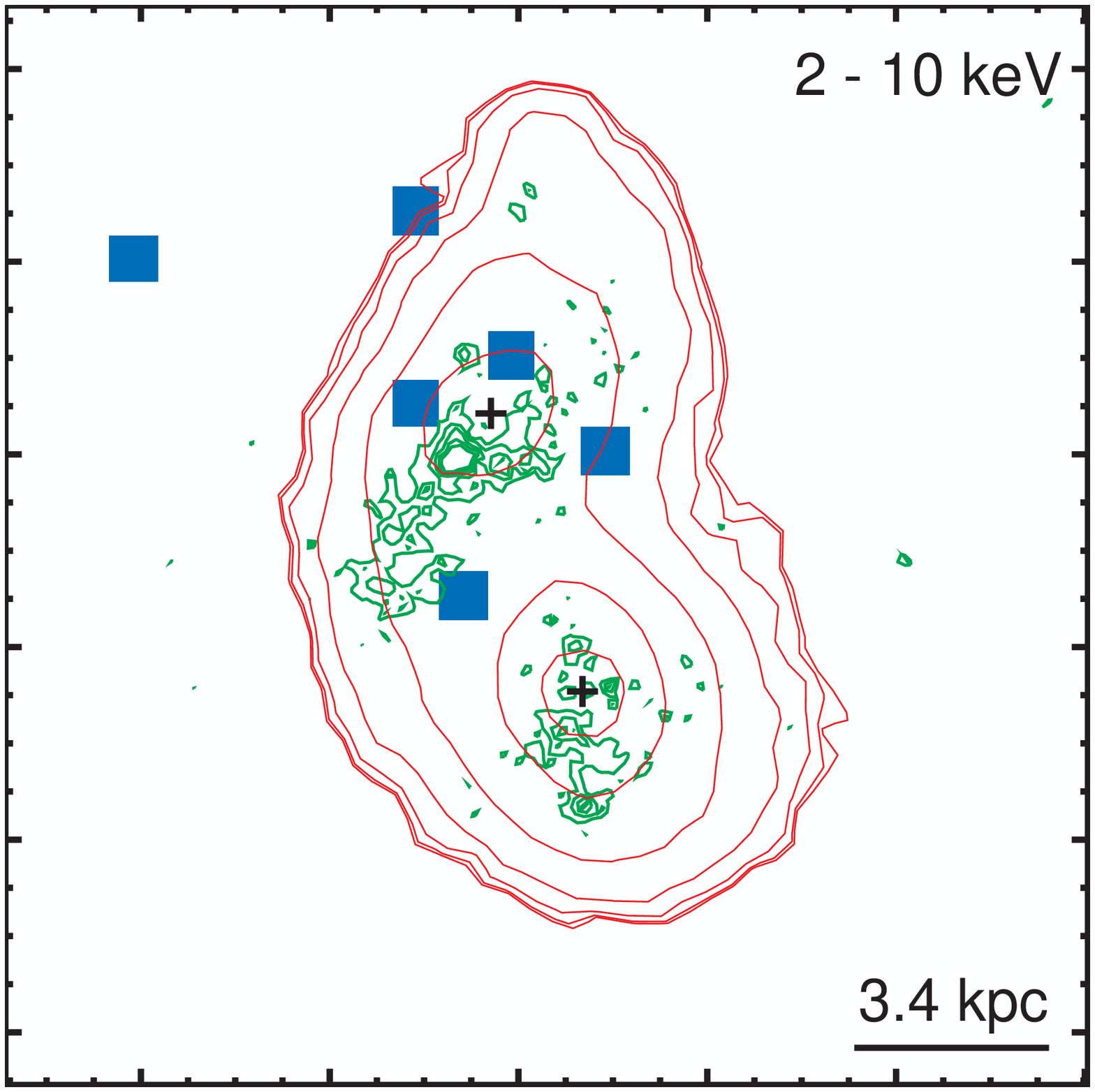}
    \caption{{\it Chandra} ACIS X-ray images ($0.''25$ pixel$^{-1}$ binning; unsmoothed)
    of the four optically selected kpc-scale binary AGNs. The left, middle, and right 
    columns
    show the full band, and zoomed-in soft and hard bands, respectively.
    Contours are {\it HST}/WFC3 F105W ($Y$ band; in red) and F336W ($U$ band; in green) images
    in linear spacing. Black crosses indicate $Y$-band nuclear positions.
    Typical absolute (relative) astrometric uncertainty is $0.''20$ ($0.''15$)
    for our ACIS images (see Section \ref{subsubsec:acis_astro} for details)
    and $0.''20$ ($0.''01$) for the $Y$-band images.
    Major tickmarks are separated by $1.''0$.}
    \label{fig:xray}
\end{figure*}

\section{Observations, Data Reduction, and Analysis}\label{sec:obs}

\subsection{{\it HST}/WFC3 F336W and F105W Imaging}\label{subsec:hst}

The four optically selected kpc-scale binary AGNs were observed
using the WFC3 on board the {\it HST} in Cycle 18 (program: GO
12363; PI: Shen). Each target was imaged in the UVIS/F336W
\citep[$U$ band;][]{dressel10} and IR/F105W (wide $Y$ band)
filters within a single {\it HST} orbit. We refer to Paper II
for details of our {\it HST} observations. Here we briefly
describe the data relevant for addressing the nature of the
ionizing sources.

The $U$- and $Y$-band images were calibrated both
photometrically and astrometrically. The typical relative
astrometric accuracy is $0.''004$ for the $U$-band and $0.''01$
for the $Y$-band images. To improve absolute astrometric
accuracy and to compare with X-ray images, we have registered
the $U$- and $Y$-band images with the SDSS astrometry. The
resulting absolute astrometric uncertainties of the registered
$U$- and $Y$-band images were estimated as $\sim 0.''25$ and
$\sim 0.''20$, respectively (Paper II). We list the $Y$-band
nuclear positions in Table \ref{table:astrometry} and the
inferred separations between the double nuclei for each target
in Table \ref{table:obs}. These $Y$-band nuclear separations
agree with those measured from our ground-based NIR imaging
within uncertainties.

At the redshifts of our targets ($z=$0.130--0.207; Table
\ref{table:obs}), the $U$-band filter covers rest-frame
$\sim2600$--3200 \angstrom . For obscured AGNs, this wavelength
range is likely to be dominated by continuum photospheric
emission from host galaxy young stellar populations, which
offers a useful indicator for star formation rate \citep[SFR;
e.g., ][]{cram98}. The $U$-band filter also covers line
emission from ionized gas, although the contamination from even
the strongest line, \MgIIb , is likely insignificant ($<$1\%),
given that its typical equivalent width is small (e.g.,
$5.2\pm0.8$ \angstrom , as measured from the composite spectrum
of Type 2 AGNs by \citealt{zakamska03}). The $U$-band images
may also contain AGN light from the obscured nuclei scattered
into our line-of-sight by dust and/or gas
\citep[e.g.,][]{zakamska06}, but our data suggest that the
contribution is likely insignificant ($<$5\%), because of the
moderate AGN luminosities of our sample (Table
\ref{table:astrometry}) and the absence of a broad \hbeta\ or
\halpha\ component in the optical spectra
\citep[e.g.,][]{liu09}. Finally, the $U$-band could also
contain nebular continua emitted by the ionized gas associated
with the AGN emission line region \citep{osterbrock89}. Based
on the observed fluxes and reddening estimates of the Balmer
lines from our optical slit spectra \citep{liu10b}, we estimate
that the nebular continua contribute to $\sim5$\%--20\% of the
$U$-band flux at $2900$ \angstrom . Therefore, the nebular
continua could make a considerable contamination to, but do not
dominate the observed $U$-band flux. The possible contamination
from all these non-star formation related processes would lower
the estimated SFRs and strengthen the case of an AGN component.

Our high-resolution $U$-band imaging provides constraints on
the intensity and spatial distribution of star formation
activity in the host galaxies (Paper II). The inferred nuclear
SFRs are useful for estimating the X-ray contribution from
star-formation-related processes (Section \ref{subsec:sf}). In
Table \ref{table:flux}, we list the observed $U$-band flux
$f_{U}$ and luminosity $L^{{\rm obs}}_{U}$ integrated within a
region centered on each $Y$-band nucleus position, with an
aperture size matched to that of X-ray extraction (Sections
\ref{subsubsec:extract} and \ref{subsubsec:decompose}). To
estimate the intrinsic $L_{U}$, we have adopted an extinction
correction for each nucleus (Table \ref{table:flux}), based on
the Balmer decrement $F_{{\rm H}\alpha}/F_{{\rm H}\beta}$
measured from our spatially resolved optical spectroscopy
\citep{liu10b}. We assume the extinction curve of
\citet{cardelli89} with $R_V = 3.1$ to calculate the Balmer
decrement. We have carefully subtracted the host galaxy stellar
continuum using population synthesis models \citep{bc03} with
the fitting method of \citet{liu09}, to avoid the ratio
$F_{{\rm H}\alpha}/F_{{\rm H}\beta}$ being overestimated due to
strong Balmer absorption in post-starburst populations. To
estimate SFR from $L_U$, we have adopted the empirical
calibration of \citet{hopkins03}, which is given by
\begin{equation}\label{eq:sfr_lu}
\frac{{\rm SFR}_{U}}{M_{\odot}~{\rm yr}^{-1}} = \bigg(\frac{L_{U}}{1.81 \times 10^{28}~
{\rm erg~s}^{-1} {\rm Hz}^{-1}}\bigg)^{1.186},
\end{equation}
with an rms scatter of $0.13$ dex. The relation was based on
the SDSS $u$-band luminosity (corrected for obscuration) of
2625 star-forming galaxies in the SDSS DR1 \citep{SDSSDR1},
which was calibrated against SFR estimates inferred from
\halpha\ emission-line measurements (corrected for aperture and
obscuration effects) according to the relation of
\citet{kennicutt98}; it is valid\footnote{For the three weakest
$U$-band nuclei in our sample (Table \ref{table:flux}), the
$L_U$ measurements are $\sim10$ times lower than the faint
luminosity end of the SDSS star-forming galaxy sample studied
by \citet{hopkins03}. To estimate SFRs for these faint nuclei,
we assume the SFR-$L_{U}$ relation by \citet{hopkins03}
extrapolated to lower luminosities.} for $2\times 10^{28}~{\rm
erg~s}^{-1} {\rm Hz}^{-1} \lesssim L_U \lesssim 10^{30}~{\rm
erg~s}^{-1} {\rm Hz}^{-1}$. \citet{hopkins03} have shown that
the inferred ${\rm SFR}_{U}$ is consistent with SFR estimates
from the 1.4 GHz luminosity \citep[which was in turn calibrated
from the FIR luminosity according to
\citealt{kennicutt98};][]{bell03} with an rms scatter of $0.23$
dex. Below in Section \ref{subsubsec:uncertainty} we discuss
systematics and uncertainties in our SFR estimates and by
extension the inferred X-ray luminosity due to star
formation-related processes (Section \ref{subsec:sf}).

\begin{deluxetable*}{lcccccccc}
\tabletypesize{\scriptsize} \tablecolumns{9}
\tablewidth{\textwidth}
%
\tablecaption{HST/WFC3 $Y$-band and Chandra/ACIS X-ray 0.2--10
keV Astrometry of the Double Nuclei\label{table:astrometry}}
\tablehead{\colhead{} & \colhead{Redshift} &
\colhead{log$L^{{\rm obs}}_{{\rm [O\,III]}}$} &
\colhead{log$L^{{\rm cor}}_{{\rm [O\,III]}}$} &
\colhead{R.A.$_{Y}$} & \colhead{Dec.$_{Y}$} &
\colhead{R.A.$_{{\rm X-ray}}$} & \colhead{Dec.$_{{\rm X-ray}}$}
& \colhead{$\Delta\theta_{{\rm diff}}$}   \\
\colhead{Object Name} & \colhead{$z_e$} & \colhead{(erg
s$^{-1}$)} & \colhead{(erg s$^{-1}$)} & \colhead{(J2000)} &
\colhead{(J2000)} & \colhead{(J2000)} & \colhead{(J2000)} &
\colhead{($''$)}  \\
\colhead{(1)} & \colhead{(2)} & \colhead{(3)} & \colhead{(4)} &
\colhead{(5)} & \colhead{(6)} & \colhead{(7)} & \colhead{(8)} &
\colhead{(9)}
}
\startdata
SDSS J1108+0659NW & 0.1812 & 42.16 & 42.66 & 11:08:51.029 & +06:59:01.32 & 11:08:51.031 &
+06:59:01.26 & 0.07  \\
SDSS J1108+0659SE & 0.1820 & 41.52 & 41.64 & 11:08:51.061 & +06:59:00.81 & 11:08:51.069 &
+06:59:00.66 & 0.19  \\
SDSS J1131$-$0204W& 0.1454 & 41.40 & 42.34 & 11:31:26.042 & $-$02:04:59.33 & \nodata &
\nodata & \nodata \\
SDSS J1131$-$0204E& 0.1470 & 41.31 & 42.22 & 11:31:26.088 & $-$02:04:59.21 & \nodata &
\nodata & \nodata \\
SDSS J1146+5110SW & 0.1293 & 41.93 & 42.18 & 11:46:42.466 & +51:10:29.46 & 11:46:42.504 &
+51:10:29.45 & 0.36 \\
SDSS J1146+5110NE & 0.1303 & 41.38 & 41.65 & 11:46:42.630 & +51:10:31.69 & 11:46:42.672 &
+51:10:32.03 & 0.52 \\
SDSS J1332+0606SW & 0.2057 & 41.06 & 41.64 & 13:32:26.340 & +06:06:27.31 & \nodata &
\nodata & \nodata \\
SDSS J1332+0606NE & 0.2074 & 41.83 & 42.84 & 13:32:26.372 & +06:06:28.73 & 13:32:26.364 &
+06:06:28.63 & 0.16 \\
\enddata
\tablecomments{Col. (2): emission-line redshift measured from
spatially resolved optical spectra \citep{liu10b}. Col. (3):
observed \OIIIb\ emission-line luminosity measured from
spatially resolved optical spectra \citep{liu10b}. Cols. (4) \&
(5): coordinates of the double nuclei measured from HST
$Y$-band images. Typical absolute (relative) astrometric
uncertainty is $0.''2$ ($0.''01$); Cols. (6) \& (7):
coordinates measured from ACIS images. Typical absolute
(relative) astrometric uncertainty is $0.''2$ ($0.''15$). See
Section \ref{subsubsec:acis_astro} for details; Cols. (8) \&
(9): difference between the X-ray and $Y$-band measured
positions.}
\end{deluxetable*}

\begin{deluxetable*}{lccccccccc}
\tabletypesize{\scriptsize} \tablecolumns{10}
\tablewidth{\textwidth}
%
\tablecaption{Nuclear $U$-band Fluxes and Luminosities,
Extinction Estimates, and Inferred Star Formation
Properties\label{table:flux}}
\tablehead{ \colhead{} & \colhead{log$f^{{\rm obs}}_{U}$} &
\colhead{log$L^{{\rm obs}}_{U}$} & \colhead{} &
\colhead{$E(B-V)$} & \colhead{$A_{U}$} & \colhead{log$L_{U}$} &
\colhead{SFR$_U$} & \colhead{L$^{{\rm SF}}_{X, 0.5-2~{\rm
keV}}$} &
\colhead{L$^{{\rm SF}}_{X, 2-10~{\rm keV}}$} \\
\colhead{Object Name} & \colhead{(Jy)} & \colhead{(erg s$^{-1}$
Hz$^{-1}$)} & \colhead{$\frac{F_{{\rm H}\alpha}}{F_{{\rm
H}\beta}}$} & \colhead{(mag)} & \colhead{(mag)} & \colhead{(erg
s$^{-1}$ Hz$^{-1}$)} & \colhead{($M_{\odot}$ yr$^{-1}$)} &
\colhead{(10$^{40}$ erg s$^{-1}$)} & \colhead{(10$^{40}$ erg s$^{-1}$)} \\
\colhead{(1)} & \colhead{(2)} & \colhead{(3)} & \colhead{(4)} &
\colhead{(5)} & \colhead{(6)} & \colhead{(7)} & \colhead{(8)} &
\colhead{(9)} & \colhead{(10)}
} \startdata
%
SDSS J1108+0659NW  & $-$4.55   & 28.4    & 4.1 & 0.4 & 1.9$\pm$0.7 & 29.2$\pm$0.3    & 10
& 6     & 6 \\
SDSS J1108+0659SE  &$<$$-$5.93 & $<$27.0 & 3.1 & 0.1 & 0.5$\pm$0.8 & $<$27.2$\pm$0.3 &
$<$0.1& $<$0.03& $<$0.03 \\
SDSS J1131$-$0204W &$<$$-$5.97 & $<$26.8 & 5.6 & 0.7 & 3.8$\pm$0.3 & $<$28.3$\pm$0.1 &
$<$1
& $<$0.5 & $<$0.6 \\
SDSS J1131$-$0204E & $-$5.69   & 27.1    & 5.5 & 0.7 & 3.7$\pm$0.4 & 28.5$\pm$0.2    & 2
& 1      & 1 \\
SDSS J1146+5110SW  & $-$4.91   & 27.7    & 3.1 & 0.1 & 0.4$\pm$0.1 & 27.9$\pm$0.1    & 0.4
& 0.2    & 0.2 \\
SDSS J1146+5110NE  & $-$4.44   & 28.2    & 3.1 & 0.1 & 0.5$\pm$0.1 & 28.4$\pm$0.1    & 2
& 0.7    & 0.8 \\
SDSS J1332+0606SW  & $-$5.39   & 27.7    & 3.5 & 0.2 & 0.9$\pm$0.3 & 28.1$\pm$0.1    & 0.6
& 0.3    & 0.3 \\
SDSS J1332+0606NE  & $-$5.25   & 27.8    & 4.7 & 0.5 & 2.3$\pm$0.9 & 28.8$\pm$0.4    & 4
& 2      & 2  \\
\enddata
\tablecomments{All measurements for each nucleus were made
within the same aperture as for the X-ray extraction. Col. (2):
observed $U$-band flux density. Col. (3): observed $U$-band
luminosity density. Col. (4): Balmer decrement measured from
our ground-based slit spectra \citep{liu10b}. Col. (5): color
excess estimated from the Balmer decrement. Col. (6): $U$-band
extinction estimated from the Balmer decrement and uncertainty
due to aperture coverage mismatch (see Section
\ref{subsubsec:uncertainty}). Col. (7): extinction-corrected
$U$-band luminosity density and uncertainty (propagated from
the extinction uncertainty due to aperture mismatch only, not
including that due to uncertain dust geometry). Col. (8): star
formation rate inferred from the extinction-corrected $U$-band
luminosity density (see Section \ref{subsec:hst} for details).
Typical uncertainty is $\sim0.3$ dex, which was estimated by
convolving that propagated from $L_U$ with the rms scatter of
the SFR-$L_U$ calibration (Equation \ref{eq:sfr_lu}). Cols. (9)
\& (10): X-ray luminosities inferred from the SFR estimate,
assuming the empirical calibration of \citet[][see Section
\ref{subsec:sf} for details]{ranalli03}. Typical uncertainty is
$\sim0.4$ dex, which was estimated by convolving that
propagated from SFR estimates with the rms scatter of the
SFR-$L_X$ calibrations (Equations \ref{eq:lxs_sfr} and
\ref{eq:lxh_sfr}).}
\end{deluxetable*}

\subsection{{\it Chandra} ACIS X-Ray Imaging Spectroscopy}\label{subsec:chandra}

The four kpc-scale binary-AGN candidates were observed with the
ACIS-S on board the {\it Chandra X-ray Observatory} between
2011 February and November (Cycle 12 program: GO1-12127X; PI:
Shen). Exposure times ranged from 16 ks to 24 ks (Table
\ref{table:xray}). They were set by the requirement of
obtaining $\sim 100$ counts in the 0.5--10 keV from the weaker
nucleus of each target. The counts were estimated from the
\OIII\ luminosity for each nucleus, using the empirical
correlation between 2--10 keV (unabsorbed) and \OIII\
luminosities (extinction corrected) from \citet{panessa06} as
the baseline value, taking into account systematic
uncertainties using the \citet{heckman05} relation (for
observed luminosities) for optically selected single Type 2
AGNs, assuming a single power-law spectrum with an absorbing
column density $N_{{\rm H}}=10^{23}\ {\rm cm^{-2}}$
\citep[typical for Type 2 Seyferts;][]{bassani99} and a photon
index $\Gamma=1.8$ \citep[typical for unabsorbed
Seyferts,][]{green09}. All the targets were observed on-axis on
the S3 chip, 9$''$ to 15$''$ away from the aimpoint. We
examined the light curves and found no flares of either the
sources or the background in each of the observations.

We reprocessed the data using the standard \chandra\
Interactive Analysis of Observations (CIAO) software
\citep{fruscione06} with version 4.4. We ran the
\textsf{chandra\_repro} script on the standard Level 2 event
file for the recommended processing steps by the Chandra X-ray
Center, applying the latest calibration files (CALDB 4.4.1).
The process corrected for charge transfer inefficiency and
time-dependent gain. The energy-dependent subpixel event
repositioning (EDSER) algorithm \citep{li04} was applied to
improve the image quality of ACIS-S data for sources near the
optical axis of the telescope, where the point spread function
(PSF) is under sampled by the $0.''5$ ACIS pixels. Our targets
have low count rates, so pileup effects \citep{ballet99} were
insignificant.

\subsubsection{Astrometric Uncertainty of ACIS
Images}\label{subsubsec:acis_astro}

First we discuss astrometry of ACIS images and our effort of
obtaining accurate alignment between the X-ray and optical
images, which are important to determining the nature of X-ray
sources. For sources within 3 arcmin of the aimpoint, the
typical absolute ACIS-S astrometric accuracy is $\lesssim
0.''6$\footnote{This was inferred based on measuring the
distances between the \chandra\ X-ray source positions and
corresponding optical/radio counterpart positions from the
Tycho2 \citep[with astrometric accuracy of $\sim25$
mas;][]{hog00} and ICRS \citep[with astrometric accuracy of
$\sim1$ mas;][]{ma98} catalogs.} (radius size of the overall
90\% uncertainty circle of ACIS-S absolute position; \chandra\
Proposers' Observatory
Guide\footnote{http://cxc.harvard.edu/proposer/POG/; see also
http://cxc.harvard.edu/cal/ASPECT/celmon/.}, hereafter POG).
The relative astrometric accuracy is $0.''15$ (90\% limit) for
on-axis sources\footnote{Based on the 900 ks ACIS-I observation
of the Orion Nebula; POG.}.

Astrometric calibration was applied as part of the pipeline
processing of ACIS images. To verify the astrometric accuracy,
we ran \textsf{wavdetect} \citep{freeman02}, which is a
wavelet-based algorithm for spatial analysis of Poisson data,
to detect sources as references for any fine alignment, if
needed, between the ACIS images and the SDSS, to which our
$Y$-band images have been registered. We used \textsf{mkpsfmap}
to create observation-specific PSF map files instead of using
the PSF table. We adopted a high significance threshold
(\textsf{sigthresh}$=$10$^{-8}$, corresponding to one spurious
source in a 10$^4\times$10$^4$ pixel map) to ensure robust
source detection.

For SDSS J1108+0659 (SDSS J1131$-$0204), the astrometry of the
ACIS image agrees with that of the SDSS within $0.''2$
($0.''2$), based on three (six) SDSS-matched sources detected
by \chandra\ within 2 arcmin (5 arcmin) of the aimpoint. For
SDSS J1146+5110, only one bright X-ray source was detected in
the field of view, 1.6 arcmin away from the aimpoint, whose
ACIS and SDSS positions agree within $0.''2$. For SDSS
J1332+0606, the astrometry of the ACIS image agrees with that
of the SDSS within $0.''3$, based on four SDSS-matched sources
detected within 3 arcmin of the aimpoint. Given these results,
we do not apply any further astrometry correction for the ACIS
images, because the agreement is already comparable to the SDSS
astrometric accuracy.

Figure \ref{fig:xray} shows the unsmoothed ACIS images of our
four targets in the full (0.5--10 keV), soft (0.5--2 keV), and
hard (2--10 keV) bands, respectively. Given the low count
levels (Table \ref{table:xray}), we do not apply any smoothing
to avoid artifacts. As we will show in Section
\ref{subsec:x2oratio}, our targets are significantly weaker
hard X-ray emitters than those predicted from both the
\citet{panessa06} and \citet{heckman05} relations for single
optically selected AGNs (by $\sim0.8\pm0.2$ dex and
$\sim1.9\pm0.3$ to $2.4\pm0.3$ dex in observed and unabsorbed
2--10 keV luminosities, respectively), resulting in far fewer
counts than we expected. Five of the eight nuclei in our
targets were detected in the full band (both nuclei in SDSS
J1108+0659, both nuclei in SDSS J1146+5110, and the NE nucleus
in SDSS J1332+0606), of which four were detected in both soft
and hard bands (both nuclei in SDSS J1108+0659, the SW nucleus
in SDSS J1146+5110, and the NE nucleus in SDSS J1332+0606),
whereas one was only detected in the soft band (the NE nucleus
in SDSS J1146+5110). The other three nuclei were undetected in
the X-rays (both nuclei in SDSS J1131-0204, and the SW nucleus
in SDSS J1332+0606).

\begin{deluxetable*}{lcccccccc}
\tabletypesize{\scriptsize} \tablecolumns{9}
\tablewidth{\textwidth}
%
\tablecaption{X-ray Properties and Spectral
Models\label{table:xray}}
\tablehead{\colhead{} & \colhead{$N_{{\rm H}}$(Galactic)} &
\colhead{Exposure} & \colhead{$T$} & \colhead{$S$} &
\colhead{$H$} &
\colhead{} & \colhead{} & \colhead{$N_{{\rm H}}$($\Gamma = 1.8$)} \\
\colhead{Object Name} & \colhead{($10^{20}$ cm$^{-2}$)} &
\colhead{(seconds)} & \colhead{(counts)} & \colhead{(counts)} &
\colhead{(counts)} & \colhead{HR} &
\colhead{$\Gamma$} & \colhead{($10^{22}$ cm$^{-2}$)} \\
\colhead{(1)} & \colhead{(2)} & \colhead{(3)} & \colhead{(4)} &
\colhead{(5)} & \colhead{(6)} & \colhead{(7)} & \colhead{(8)} &
\colhead{(9)}
}
\startdata
%
SDSS J1108+0659NW  & 3.99 & 19273 & 25.4$\pm$5.0 & 20.5$\pm$4.5 & 4.3$\pm$2.1 &
$-0.60^{+0.11}_{-0.19}$ & $2.4^{+0.6}_{-0.4}$ & $<$10$^{-2}$ \\
SDSS J1108+0659SE  & 3.99 & 19273 & 12.4$\pm$3.5 & 2.5$\pm$1.6  & 9.1$\pm$3.0 &
$0.46^{+0.32}_{-0.16}$ & $0.0^{+0.4}_{-0.6}$ & 3$^{+3}_{-1}$ \\
SDSS J1131$-$0204W & 3.34 & 23723 & $<$6.6   & \nodata     & \nodata      & \nodata &
\nodata & \nodata \\
SDSS J1131$-$0204E & 3.34 & 23723 & $<$6.6   & \nodata     & \nodata      & \nodata &
\nodata & \nodata \\
SDSS J1146+5110SW & 1.58 & 15432 & 16.6$\pm$5.2& 2.9$\pm$2.9 & 13.6$\pm$4.8
&$0.65^{+0.19}_{-0.17}$ & $-0.4^{+0.4}_{-0.5}$ & 4.0$\pm$1.5 \\
SDSS J1146+5110NE & 1.58 & 15432 & 4.6$\pm$3.4 & 4.6$\pm$3.4 & $<$1.8       & $<$$-0.81$ &
$>$3.0          & $<$10$^{-2}$ \\
SDSS J1332+0606SW & 2.21 & 23952 & $<$10.9       & \nodata     & \nodata      & \nodata &
\nodata & \nodata \\
SDSS J1332+0606NE & 2.21 & 23952 & 7.7$\pm$2.9 & 3.9$\pm$3.2 & 3.7$\pm$3.2  &
$-0.03^{+0.31}_{-0.36}$ & $1.1^{+0.8}_{-0.7}$ & $1.0^{+2.0}_{-0.99}$ \\
\enddata
\tablecomments{Col. (2): Galactic column density, calculated
adopting the neutral hydrogen data set compiled by
\citet{dickey90}, using the CIAO observing toolkit at
http://asc.harvard.edu/toolkit/colden.jsp. Col. (3): ACIS
exposure time. Col. (4): total 0.5--10 keV counts. Col. (5):
soft 0.5--2 keV counts. Col. (6): hard 2--10 keV counts. Col.
(7): Hardness ratio, HR$\equiv(H-S)/(H+S)$. Col. (8): photon
index of a power-law model where $n(E)\propto E^{-\Gamma}$,
assuming Galactic column density. Col. (9): Intrinsic column
density estimated assuming a power-law model with $\Gamma=1.8$,
a typical value for unobscured AGNs \citep[e.g.,][]{green09},
absorbed by a gas column $N_{{\rm H}}$.}
\end{deluxetable*}

Of the three X-ray detected targets, the $Y$-band nuclei in
SDSS J1146+5110 and in SDSS J1332+0606 are separated by $1.''5$
and $2.''7$, respectively (Table \ref{table:obs}), which are
well within the resolving power of ACIS; the $Y$-band nuclei in
SDSS J1108+0659 require more careful decomposition (Section
\ref{subsubsec:decompose}). For the X-ray detected nuclei in
SDSS J1146+5110 and SDSS J1332+0606, we compare their positions
with the $Y$-band nuclear positions. We measure the X-ray
position with a two-dimensional (2D) image fitting analysis
using \textsf{Sherpa} \citep{freeman01}. We adopt a constant
for the background and the PSF images as convolution kernels to
fit the sources. We applied the $Y$-band positions as the
initial guesses. PSF images were created with the \textsf{MARX}
software\footnote{http://space.mit.edu/ASC/MARX/index.html}
using the PSF-ray table generated by the \chandra\ Ray Tracer
\citep[ChaRT;][]{carter03}, which simulates the best available
PSF at any off-axis angle and for any energy or spectrum. As
listed in Table \ref{table:astrometry}, the X-ray positions
agree with the $Y$-band positions within the uncertainties for
all the detected sources. We have compared the radial profile
of our targets against PSF models. Each nucleus component is
consistent with being an unresolved point source.

\subsubsection{Source Extraction for Well-separated Nuclei}\label{subsubsec:extract}


For SDSS J1146+5110 and SDSS J1332+0606, we run {\it dmextract}
to extract X-ray counts for each individual nucleus. We use the
positions and their uncertainties measured for each nucleus
from the {\it HST} $Y$-band images as priors for source
extraction. The background counts were extracted from
source-free regions around the target regions. We report the
number of background subtracted counts in separate soft (0.5--2
keV) and hard (2--10 keV) band and in the full band in Table
\ref{table:xray}.

For each of the two nuclei in SDSS J1146+5110, we adopted a
circular region with a $1.''2$ radius for source extraction.
The adopted size ensured no overlap between the two sources,
with each region containing more than 95\% of the encircled
energy (two dimensional integral of the PSF). For SDSS
J1332+0606, we extracted the counts for each nucleus in
circular regions of $1.''0$ radii, to ensure that there was no
overlap between the extraction regions of the two nuclei. No
significant detection was obtained for the southern nucleus in
SDSS J1332+0606; we estimated a 3-$\sigma$ upper limit
according to the inferred background counts level using the
tables in \citet{gehrels86} appropriate for small numbers of
events.

\begin{figure*}
  \centering
    \fbox{\includegraphics[width=88mm]{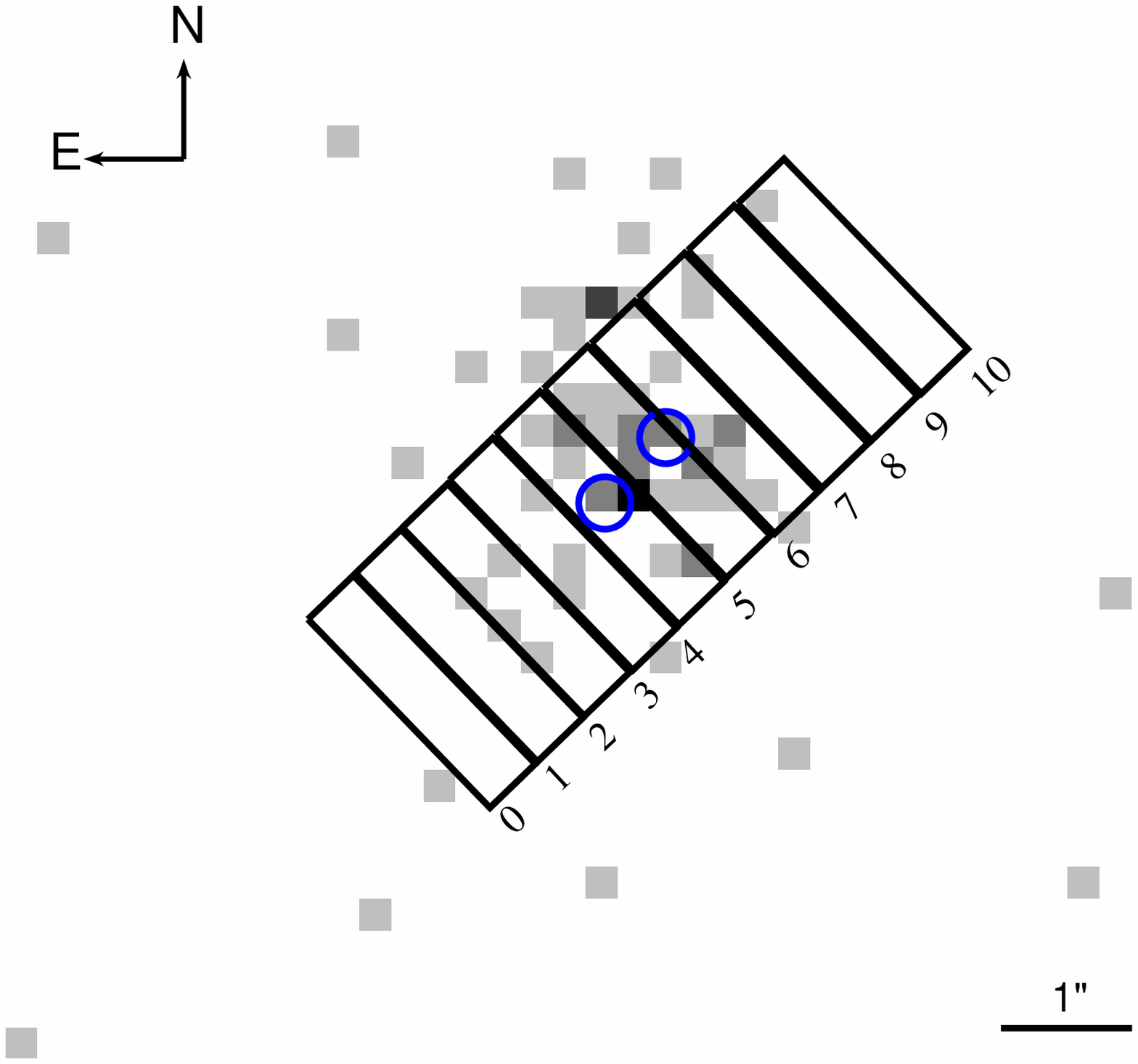}}
    \includegraphics[width=88mm]{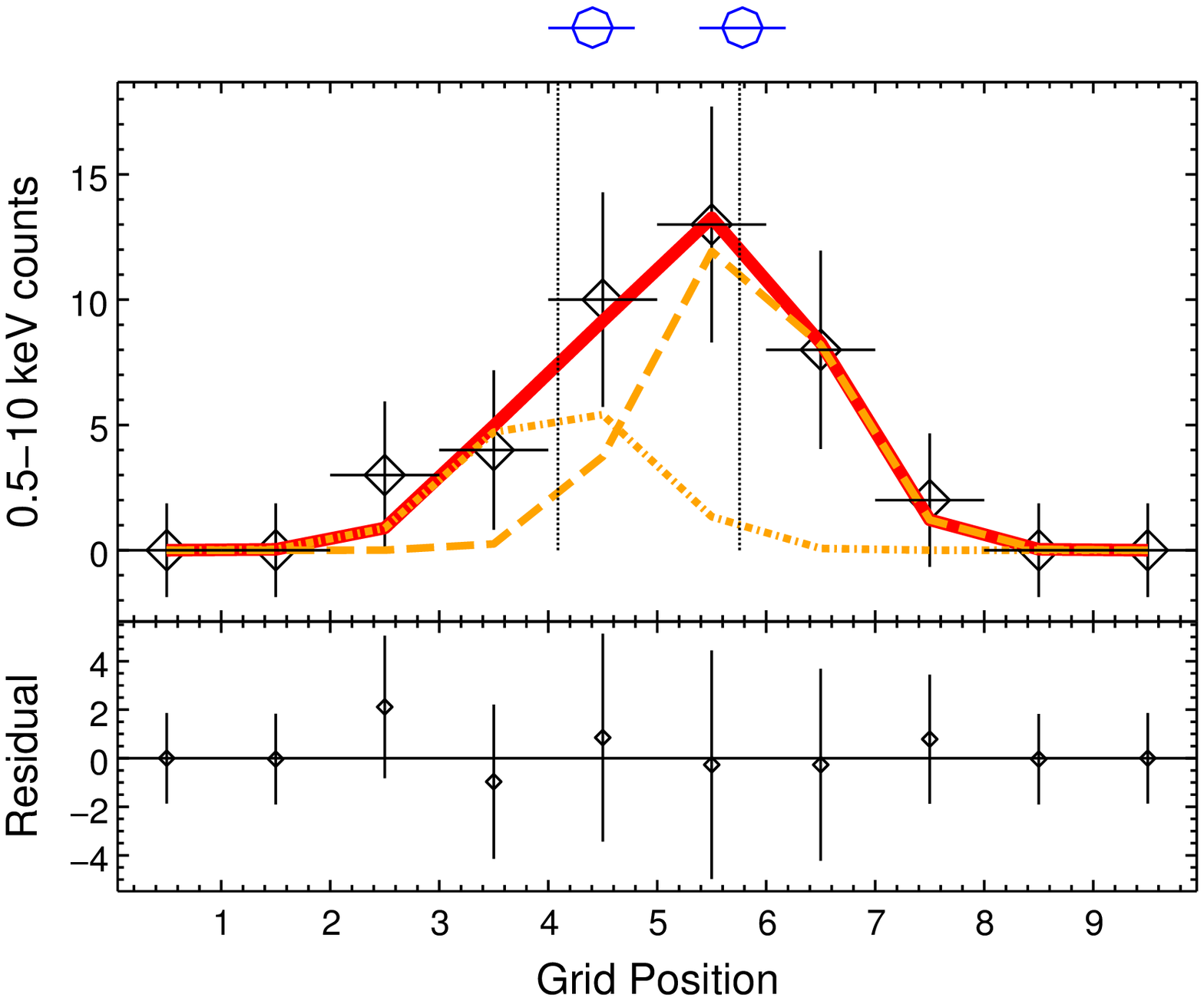}
    \caption{Left panel: the grid region for constructing the 1D spatial
    profile for SDSS J1108+0659. Background is the {\it Chandra} ACIS X-ray full-band
    image (unsmoothed with $0.''25$ pixel$^{-1}$ binning). The blue circles indicate
    the $Y$-band positions and uncertainties of the double stellar nuclei
    from our {\it HST} WFC3 imaging.
    Right panel: projected 1D profile of the
    nuclear X-ray emission in the full band. The grid positions correspond to
    those as shown by the grid region in the left panel and are separated by $0.''5$
    (major tick marks).
    Data are shown in diamonds with error bars, whereas our best-fit models
    are displayed in solid (for the total) and dotted/dashed
    (for each individual component) curves, respectively.
    The vertical lines indicate the X-ray centers of the two components. The blue circles
    on top indicate the projected $Y$-band positions of the double stellar nuclei.}
    \label{fig:1108onedfit}
\end{figure*}

\begin{figure*}
  \centering
    \includegraphics[width=88mm]{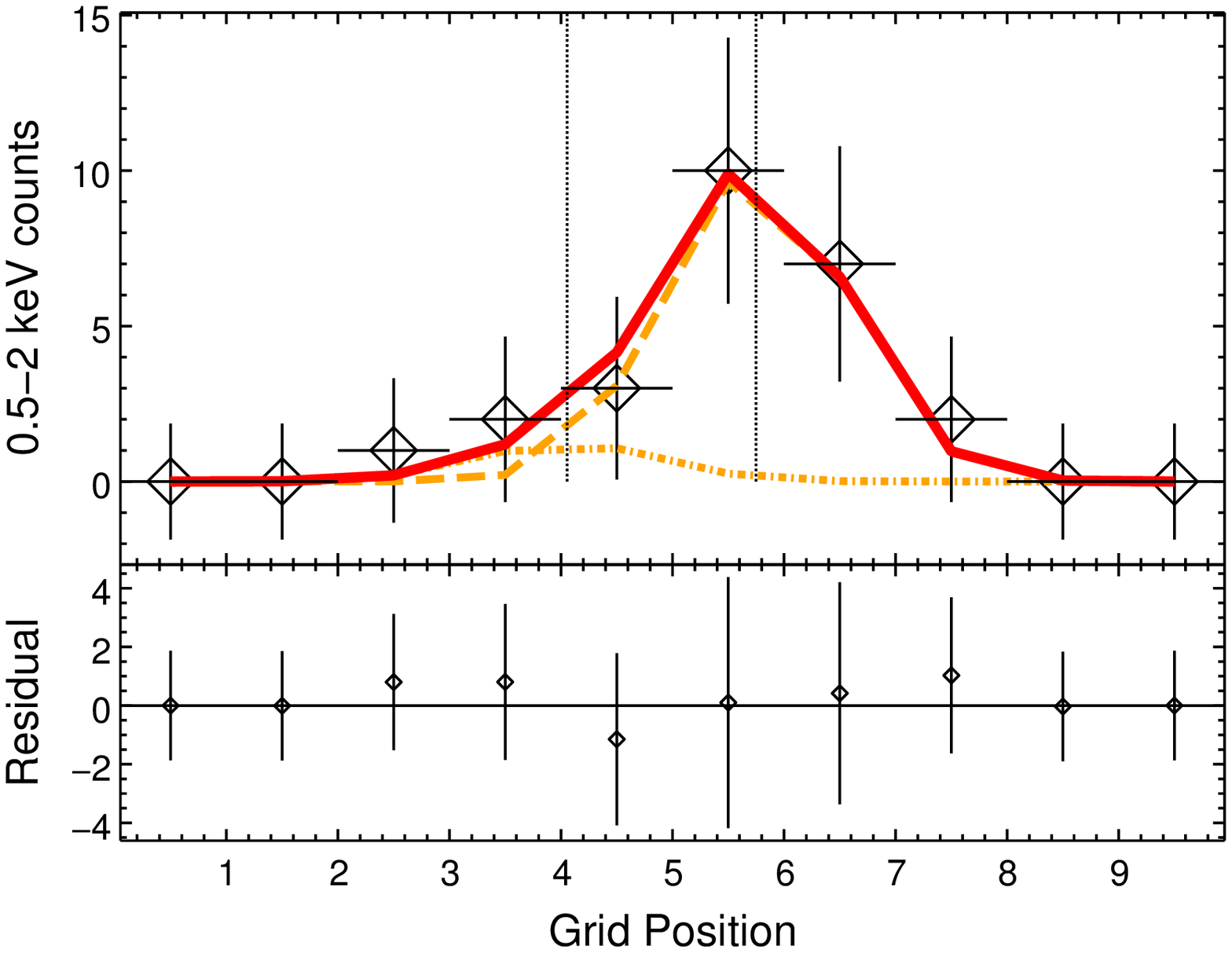}
    \includegraphics[width=88mm]{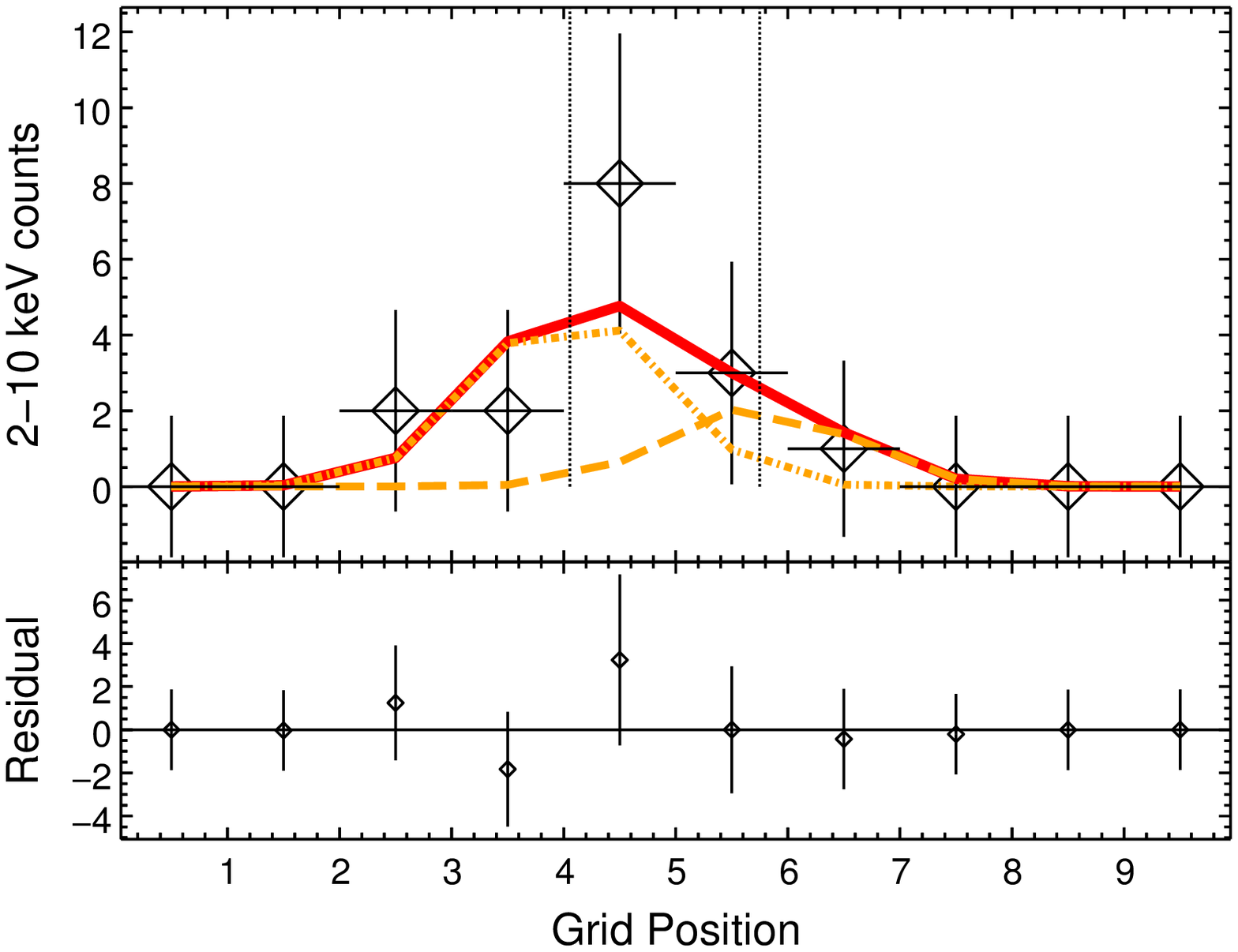}
    \caption{Same as in the right panel of Figure \ref{fig:1108onedfit},
    but for the soft (left panel) and hard (right panel) band, respectively.}
    \label{fig:1108spec}
\end{figure*}

\subsubsection{Source Decomposition for Marginally Resolved
Nuclei}\label{subsubsec:decompose}

The $Y$-band nuclei of SDSS J1108+0659 and SDSS J1131$-$0204
are both separated by $0.''70$ (Table \ref{table:obs}), which
is close to the limit of the resolving power of ACIS. Neither
one of the nuclei in SDSS J1131$-$0204 was detected in our
observations, and we estimated a 3 $\sigma$ upper limit using
the tables in \citet{gehrels86} at the $Y$-band position of
each nucleus.

To decompose the X-ray emission from the double nuclei in SDSS
J1108+0659, we first performed a 2D image fitting in
\textsf{Sherpa}, using PSF models as convolution kernels. We
were unable to unambiguously separate the double nuclei, whose
counts were too few for a statistically significant 2D
decomposition. To increase the S/N, we then performed a
one-dimensional (1D) analysis for the nuclear X-ray emission in
SDSS J1108+0659, following the method used in \citet{civano12}.
There were too few counts to do a similar 1D analysis in SDSS
J1131$-$0204.

For the 1D analysis in SDSS J1108+0659, we projected the X-ray
(full band) source profile in a direction connecting the two
$Y$-band nuclei by extracting the counts in a grid of regions
of $0.''5\times2.''0$, as shown in Figure
\ref{fig:1108onedfit}. The grid was designed to minimize
contamination from the extended emission to the northeast of
the nuclear region. One caveat is that the apertures used to
derive the 1D profile may still contain emission from the
extended structure, but we proceed by assuming that the
emission within the nuclear apertures can be modeled as coming
from a combination of point sources. We defer to Paper II a
full treatment of the properties and origins \citep[e.g.,
starburst, shock heated gas, scattered AGN light, outflows
and/or gas clouds photoionized by the AGN;][]{young01,evans06}
of the northeast extended soft X-ray emission. Figure
\ref{fig:1108onedfit} shows the projected 1D source profile. We
compared the 1D profile with the 2D PSF model projected and
convolved with the same grid. We performed fitting in
\textsf{Sherpa} using $\chi^2$ statistics with the Gehrels
variance function \citep{gehrels86}; we tested two scenarios
using one-component and two-component (i.e., two AGNs) PSF
models, respectively. Centroids and amplitudes of the PSF
models have been left free to vary. A likelihood ratio test
suggests that the data favor a two-component marginally over a
one-component model at the $\sim1.5$ $\sigma$ significance
level.

We show in Figure \ref{fig:1108onedfit} our best-fit model for
the 1D nuclear source profile of SDSS J1108+0659. The best-fit
central positions of the two sources are consistent (within 1
$\sigma$ uncertainties, $0.''2$) with the $Y$-band positions
projected on the axis connecting the two nuclei. We report the
best-fit X-ray nuclear positions of SDSS J1108+0659 in Table
\ref{table:astrometry}. We then repeated the decomposition
analysis for the soft and hard bands, respectively. With fewer
counts in the individual band fittings, we fixed the centroids
of the two models at the best-fit positions from the full-band
analysis, allowing only their amplitudes to vary. Figure
\ref{fig:1108spec} displays the best-fit models for the soft
and hard band, respectively. The NW component is stronger
(weaker) than the SE component in the soft (hard) band. The
spectral properties of the two sources seem to be significantly
different from each other, lending further support to the
two-component scenario. In Table \ref{table:xray}, we report
the number of background subtracted counts separately in the
soft and hard bands as well as in the full band.

\subsubsection{X-Ray Hardness Ratio and Spectral
Analysis}\label{subsubsec:xspec}

There are too few X-ray counts of our targets to perform
reliable spectral fitting. Instead, we estimate parameters of
X-ray spectral models using hardness ratios (HR) as a proxy for
detailed spectral fitting. The hardness ratio is defined as
\begin{equation}
{\rm HR} \equiv \frac{H-S}{H+S},
\end{equation}
where $H$ and $S$ are the number of counts in the hard and soft
bands, respectively. We adopted the Bayesian estimation of
hardness ratios \citep[][]{park06} to measure the HRs and their
uncertainties, appropriate for the low-count regime. To
estimate the photon index $\Gamma$, we assume a single
power-law model $n(E)\propto E^{-\Gamma}$, absorbed by the
Galactic column density, calculated using the CIAO observing
toolkit\footnote{http://asc.harvard.edu/toolkit/colden.jsp.}
based on the neutral hydrogen data set compiled by
\citet{dickey90}. We also estimate the intrinsic galactic
column density by fixing $\Gamma = 1.8$, typical for
low-redshift unobscured Seyferts
\citep[e.g.,][]{nandra94,green09}. We list in Table
\ref{table:xray} our X-ray count measurements and estimates for
model parameters for each detected nucleus.

We caution that the adopted single absorbed power-law model is
most likely too simple for the X-ray spectra of obscured AGNs,
in which thermal emission from starburst components and
scattered nuclear emission are often present
\citep[e.g.,][]{turner97,turner97b}. However, the low counts of
our detections do not allow us to test more realistic models.
In addition, our estimates of the intrinsic absorbing column
may not necessarily reflect the true values in cases of patchy
obscuration and/or significant scattering off an ionized medium
in Compton-thick (i.e., $N_{{\rm H}}\sim10^{24}$ cm$^{-1}$ or
larger) AGNs \citep[which represent about half of the local
Type 2 Seyfert population;][]{risaliti99}, as observed in NGC
6240 \citep[e.g.,][]{vignati99,ptak03} and in NGC 1068
\citep[e.g.,][]{matt97,guainazzi99}, although, again, the
quality of our data do not allow us to robustly test these
possibilities.

\section{Results}\label{sec:result}

We examine the nature of the ionizing sources in the four
optically selected kpc-scale binary AGNs. We first discuss the
intrinsic X-ray luminosity and spectral properties (Section
\ref{subsec:xraylumi}). We then estimate the contribution from
star-formation-related processes in the nuclear region to the
observed X-ray flux (Section \ref{subsec:sf}). Finally, we
address whether these new X-ray and {\it HST} observations
support the binary-AGN scenario for each target (Section
\ref{subsec:nature}), estimate the X-ray-to-\OIII\ luminosity
ratio of optical kpc-scale binary AGNs, and compare with single
AGNs (Section \ref{subsec:x2oratio}).


\begin{deluxetable*}{lccccccc}
\tabletypesize{\scriptsize} \tablecolumns{7}
\tablewidth{\textwidth}
%
\tablecaption{X-Ray Fluxes and Luminosities
\label{table:xrayflux}}
\tablehead{\colhead{} & \colhead{F$_{X,0.5-10{\rm keV}}$} &
\colhead{F$_{X,0.5-2{\rm keV}}$} &\colhead{ F$_{X,2-10{\rm
keV}}$} & \colhead{L$_{X,0.5-10{\rm keV}}$} &
\colhead{L$_{X,0.5-2{\rm keV}}$} & \colhead{L$_{X,2-10{\rm
keV}}$} \\
\colhead{Object Name} & \colhead{(10$^{-14}$ erg s$^{-1}$
cm$^{-2}$)} &\colhead{(10$^{-14}$ erg s$^{-1}$ cm$^{-2}$)}
&\colhead{(10$^{-14}$ erg s$^{-1}$ cm$^{-2}$)} &
\colhead{(10$^{42}$ erg s$^{-1}$)} &
\colhead{(10$^{42}$ erg s$^{-1}$)} & \colhead{(10$^{42}$ erg s$^{-1}$)} \\
\colhead{(1)} & \colhead{(2)} & \colhead{(3)} & \colhead{(4)} &
\colhead{(5)} & \colhead{(6)} & \colhead{(7)}
} \startdata
%
SDSS J1108+0659NW              & 1.05$\pm$0.15 & 0.41$\pm$0.08   & 0.46$\pm$0.23 &
0.91$\pm$0.17 & 0.34$\pm$0.06   & 0.40$\pm$0.20 \\
SDSS J1108+0659SE (observed)   & 1.20$\pm$0.35 &
0.054$\pm$0.030 & 1.11$\pm$0.36 & 1.08$\pm$0.31
& 0.049$\pm$0.032 & 1.00$\pm$0.29  \\
SDSS J1108+0659SE (unabsorbed) & 2.10$\pm$0.48 & 0.59$\pm$0.38
& 1.31$\pm$0.43 & 1.89$\pm$0.53
& 0.53$\pm$0.35   & 1.18$\pm$0.42  \\
SDSS J1131$-$0204W& $<$0.23 & \nodata & \nodata & $<$0.1 & \nodata& \nodata   \\
SDSS J1131$-$0204E& $<$0.23 & \nodata & \nodata & $<$0.1 & \nodata& \nodata   \\
SDSS J1146+5110SW (observed)   & 2.49$\pm$0.78 & 0.08$\pm$0.07 & 2.29$\pm$0.81 &
1.05$\pm$0.33  & 0.030$\pm$0.025 & 0.96$\pm$0.34 \\
SDSS J1146+5110SW (unabsorbed) & 5.2$\pm$1.6 & 2.18$\pm$0.77   & 3.2$\pm$1.1 &
2.20$\pm$0.69  & 0.92$\pm$0.28  & 1.30$\pm$0.46 \\
SDSS J1146+5110NE & 0.23$\pm$0.17 & 0.23$\pm$0.17 & $<$0.9 & 0.10$\pm$0.07 & 0.10$\pm$0.07
& $<$0.30  \\
SDSS J1332+0606SW & $<$0.35 & \nodata & \nodata & $<$0.37 & \nodata & \nodata   \\
SDSS J1332+0606NE (observed)   & 0.39$\pm$0.14 & 0.062$\pm$0.050 & 0.33$\pm$0.28 &
0.43$\pm$0.16 & 0.069$\pm$0.057 & 0.37$\pm$0.30 \\
SDSS J1332+0606NE (unabsorbed) & 0.56$\pm$0.21 & 0.21$\pm$0.16   & 0.35$\pm$0.30 &
0.65$\pm$0.24 & 0.23$\pm$0.19   & 0.39$\pm$0.31 \\
\enddata
\tablecomments{Cols. (2)--(4): total, soft, and hard X-ray flux
and 1 $\sigma$ error. Col. (5)--(7): total, soft, and hard
X-ray luminosity and 1 $\sigma$ error. For obscured sources,
both observed (obs) and unabsorbed (unabs) estimates are
listed. The unabsorbed estimates were calculated assuming a
power-law model with $\Gamma=1.8$ and $N_{\rm H}$ as derived
from the HR measurements, as listed in Table \ref{table:xray}.
All errors quoted are statistical uncertainties only.}
\end{deluxetable*}

\subsection{X-Ray Luminosity and Spectral Properties}\label{subsec:xraylumi}

X-ray emission provides the most direct evidence for nuclear
activity. In particular, the 2--10 keV hard X-ray band is
transparent to column densities of $N_{{\rm H}}\lesssim10^{24}$
cm$^{-2}$. To infer X-ray luminosity, we assume a simple
absorbed power-law model, with a fixed $\Gamma = 1.8$ and the
estimated intrinsic host galaxy column density as given in
Table \ref{table:xray}. In Table \ref{table:xrayflux}, we list
the observed X-ray flux and luminosity as well as the estimated
unabsorbed/intrinsic X-ray luminosity of each nucleus in the
total, soft, and hard bands, respectively.

The nuclei of our targets are optically classified as Type 2
AGNs \citep{liu10b}, whose observed \OIII\ luminosities (Table
\ref{table:astrometry}) suggest moderate AGN luminosities. The
four hard X-ray detected nuclei have estimated unabsorbed 2--10
keV luminosities ranging from $(3.9\pm3.1)\times 10^{41}$ erg
s$^{-1}$ to $(1.3\pm0.5) \times 10^{42}$ erg s$^{-1}$, and
unabsorbed 0.5--10 keV luminosities ranging from
$(6.5\pm2.4)\times 10^{41}$ erg s$^{-1}$ to $(2.2\pm0.7) \times
10^{42}$ erg s$^{-1}$. The estimated upper limits for the four
hard X-ray undetected nuclei range from $\sim1.0$ to $3.7\times
10^{41}$ erg s$^{-1}$ in 0.5--10 keV. These luminosity
estimates are similar to the few previously known X-ray
confirmed kpc-scale binary AGNs (NGC 6240, \citealt{komossa03};
3C 75, \citealt{hudson06}; Mrk 463, \citealt{bianchi08}; Mrk
266, \citealt{brassington07}; and Mrk 739, \citealt{koss11};
see also \citealt{ballo04} for a candidate, Arp 299). We devote
the rest of the section to determining the nature of the X-ray
sources.

First, we consider the possibility of ultra-luminous X-ray
sources \citep[ULXs; with typical X-ray luminosities of order
$10^{39}$--$10^{40}$ erg s$^{-1}$, i.e., beyond high-mass X-ray
binaries but much less than typical
AGNs;][]{long83,fabbiano06}, which are off-nuclear point-like
X-ray sources commonly observed in local major mergers of disk
galaxies such as in Arp 244 (i.e., the Antennae) and Arp 270
\citep[e.g.,][]{brassington07}. Some ULXs may be accreting
intermediate mass BHs \citep[$M_{{\rm BH}}\sim10^2$--$10^4
M_{\odot}$; e.g.,][]{miller03,miller04}. Since the nuclei in
our targets have X-ray luminosities much higher than typical
ULXs, they are most likely of a different origin. Further
evidence against the ULX scenario includes: (1) the BHs are
expected to be supermassive given their host bulge properties
\citep{liu10b}; (2) the observed \OIII\ luminosities are
significantly higher than those for ULXs \citep{abolmasov07};
and (3) the X-ray point sources in our targets are nuclear
given astrometric uncertainties, although ULXs could also live
close to nuclear regions.

Second, the estimated intrinsic hard X-ray luminosities of our
targets are close to or below $\sim10^{42}$ erg s$^{-1}$ -- the
characteristic upper limit for the most luminous star-forming
galaxies \citep[e.g.,][]{zezas01}. So it is quite possible that
much or all of the luminosity is due to star formation. While
X-ray spectral shape offers another diagnostic to discriminate
between AGN and starburst scenarios, the uncertainties of our
spectral estimates are too large to draw firm conclusions for
the majority of the nuclei. Therefore we must factor in some
independent SFR estimates to critically test the AGN scenario
for each nucleus.

\begin{figure*}
  \centering
    \includegraphics[width=185mm]{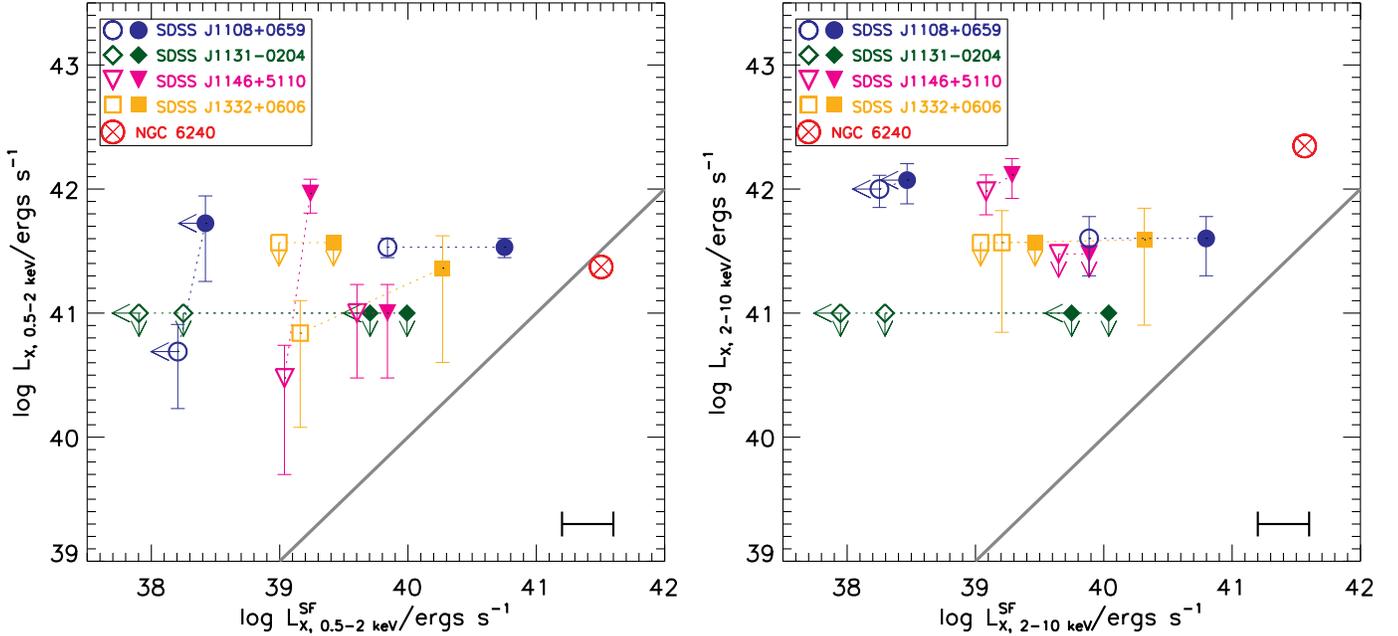}
    \caption{X-ray luminosities vs. the expected contribution from
    star-formation-related processes. We show the soft-band results in the left panel
    and the hard-band results in the right panel, respectively.
    Open symbols represent observed quantities, whereas filled symbols are
    corrected for dust extinction and gas absorption. We use dotted lines to
    connect values of the same nucleus before and after extinction/absorption
    correction. For comparison we also show measurements of NGC 6240
    (total emission from both nuclei) from the literature \citep{ptak03}.
    In both panels, we show the equality relation with a solid line.
    Error bar in the lower right of the panel indicates
    typical uncertainty ($\sim0.4$ dex) in the X-ray luminosity
    from star-formation-related processes, which was estimated by convolving that
    propagated from SFR estimates with the rms scatter of the
    SFR--$L_X$ calibrations.}
    \label{fig:lxlxsf}
\end{figure*}

\subsection{Contribution from Nuclear Star Formation}\label{subsec:sf}

Even without the presence of an AGN, star-formation-related
processes \citep[e.g., accretion onto a neutron star or a black
hole in X-ray binaries, thermal bremsstrahlung from a
starburst-driven wind;][]{strickland00} may produce strong soft
and hard X-ray emission. Intense nuclear star formation often
accompanies and sometimes outshines AGN in X-rays in gas-rich
mergers, making kpc-scale binary AGNs challenging to pin down.
To break the degeneracy, we first examine the contribution from
nuclear star formation, using independent constraints from {\it
HST} $U$-band imaging.

We estimate the expected X-ray emission due to star formation
within the same apertures used to perform our X-ray extraction,
to evaluate if an additional ionizing source, i.e., an AGN
component, is needed. To derive X-ray luminosities from SFRs,
we adopt the empirical calibration of \citet[][see also
\citealt{grimm03}]{ranalli03} based on 23 nearby star-forming
galaxies, which is given by
\begin{equation}\label{eq:lxs_sfr}
L^{{\rm SF}}_{0.5-2\, {\rm keV}} = 4.5 \times 10^{39} \frac{{\rm SFR}}{M_{\odot}~{\rm
yr}^{-1}} {\rm erg~s}^{-1}, \\
\end{equation}
\begin{equation}\label{eq:lxh_sfr}
L^{{\rm SF}}_{2-10\, {\rm keV}} = 5.0 \times 10^{39} \frac{{\rm SFR}}{M_{\odot}~{\rm
yr}^{-1}} {\rm erg~s}^{-1},
\end{equation}
with an rms scatter of 0.27 dex and 0.29 dex, respectively. In
Table \ref{table:flux}, we list the derived $L^{{\rm
SF}}_{0.5-2\, {\rm keV}}$ and $L^{{\rm SF}}_{2-10\, {\rm keV}}$
estimates for each nucleus. The predicted X-ray contribution
from star formation is an order of magnitude or more below the
observed X-ray luminosity. At this point, the case that the
X-rays are coming from AGN seems unambiguous, before
considering the uncertainties.

\subsubsection{Uncertainties}\label{subsubsec:uncertainty}

In this section, we discuss systematics and uncertainties of
our estimates of the expected X-ray luminosity due to
star-formation-related processes.  First, a Salpeter initial
mass function \citep[IMF;][]{salpeter55} was assumed with mass
limits of 0.1 $M_{\odot}$ and 100 $M_{\odot}$ in the adopted
calibration of SFR from $U$-band luminosity. The adopted
SFR--X-ray-luminosity relation of \citet{ranalli03} was
calibrated under the same assumptions about the IMF and mass
range (essentially all from \citealt{kennicutt98}), so that the
inferred star-formation-related X-ray luminosity is not
sensitive to the IMF uncertainty, given that there is no large
systematic IMF variations among star-forming galaxies
\citep{scalo86,kroupa01,chabrier03}.

The uncertainty of the estimated X-ray luminosity due to star
formation is likely to be dominated by the poorly constrained
$U$-band extinction correction. First, we estimated $F_{{\rm
H}\alpha}/F_{{\rm H}\beta}$ using emission-line measurements
for each nucleus from our ground-based optical long-slit
spectroscopy. Due to its lower angular resolution and
projection effects, the aperture of the emission-line
measurement does not exactly match with the $U$-band
measurement. To estimate the reddening uncertainty due to this
coverage mismatch, we have compared the emission-line ratio
measurement from our slit spectroscopy against that from the
SDSS fiber spectra. For the SDSS fiber-integrated measurements,
we assumed that the two velocity components correspond to the
two nucleus components, although the actual association is
likely to be more complicated. Nevertheless, we found that the
uncertainty in $U$-band extinction estimates due to aperture
mismatch is in general $<1$ mag (Table \ref{table:flux}).
Multi-color imaging and/or integral field unit spectroscopy
with the same angular resolution as the $U$-band imaging would
help further constrain the uncertainty.

More importantly, the color excesses we derived using the
Balmer decrement method do not necessarily represent the true
dust content of our targets. For example, most of the dust
could be concentrated on scales smaller than where the Balmer
lines are emitted. High extinction with little reddening could
arise in objects with very patchy and optically thick dust
clouds. These caveats associated with the uncertainties in the
extinction geometry and the reddening assumptions are typical,
but must be kept in mind when extinction corrections are
applied to compare with other studies in the literature.

\subsection{Nature of the Ionizing Sources}\label{subsec:nature}

In Figure \ref{fig:lxlxsf}, we compare the expected X-ray
luminosities due to star formation against the observed X-ray
luminosities in the soft and hard bands, respectively. Armed
with both the observed X-ray properties and independent
constraints on the expected X-ray contribution due to star
formation from $U$-band imaging, we now discuss whether the
observations support the binary-AGN scenario for each of our
targets.

\subsubsection{SDSS J1108+0659}

Both nuclei were optically classified as Type 2 Seyferts,
suggesting the presence of at least one AGN component. Both
nuclei were detected in both soft and hard X-ray bands. The
$U$-band image reveals intense star formation activity in the
NW nuclear region. While the X-ray HR and $\Gamma$ estimates
(based on the simple absorbed power-law model) suggest no
nuclear obscuration, the strong starburst component may be
accompanied by significant dust and gas, suggesting the
presence of a substantial absorbing column. The adopted single
absorbed power-law model is most likely too simple, but there
are too few X-ray counts to test more realistic multi-component
spectral models. The spectral properties of the SE source
suggest moderate nuclear obscuration, with an estimated column
density $N_{\rm H} \sim 3_{-1}^{+3}\times$10$^{22}$cm$^{-2}$.
The SE nucleus was undetected in the $U$ band, indicating very
low level of star formation around the SE nucleus (or that the
star formation was highly obscured). The spatial profiles of
the two nuclear X-ray sources support the AGN scenarios for
both, although the starburst components in the NW nucleus are
too compact to be resolved in the X-rays. For both nuclei, the
expected star-formation-related X-ray luminosities are too low
($5.7\pm1.0$ times fainter in the soft and $6.7\pm3.3$ times
fainter in hard X-ray band for the NW nucleus, and at least
three orders of magnitude fainter in both bands for the SE
nucleus) to explain the observed values, lending further
support to the binary-AGN scenario.

\subsubsection{SDSS J1131$-$0204}

Both nuclei were optically classified as Type 2 Seyferts,
indicating that at least one AGN component must be present.
Both nuclei were undetected in the two X-ray bands. The merger
system seems to be embedded in a massive disk component, and we
may be viewing the disk at an angle close to edge-on through a
large amount of absorbing column. Our $U$-band image reveals
circumnuclear star formation around the eastern nucleus,
whereas the western nucleus was undetected in the $U$-band. The
$U$-band constraints on the expected star-formation-related
X-ray luminosities are still roughly an order of magnitude
smaller than the observed X-ray luminosity upper limits for
both nuclei. Therefore, the upper limits of the X-ray
luminosities are still consistent with the presence of double
AGNs, although the possibility that one AGN ionizes gas in both
galaxies cannot be ruled out.

\subsubsection{SDSS J1146+5110}

The SW nucleus was optically classified as a Type 2 Seyfert,
whereas the NE nucleus is either a Type 2 Seyfert, a LINER, or
a LINER-H {\tiny II} composite. The X-ray spatial profiles of
the two nuclei support the AGN scenario for both, although both
nuclear starburst components are too compact to be resolved in
the X-rays. The SW nucleus was detected in both soft and hard
X-rays. Its HR and $\Gamma$ measurements may suggest a mild
obscuration ($N_{\rm H} \sim 4.0\pm1.5 \times 10^{22}$
cm$^{-2}$ estimated assuming $\Gamma$=1.8); its intrinsic hard
X-ray luminosity ($L_{X,2-10\, {\rm keV}} \sim 1.3\pm0.5 \times
10^{42}$ \lum ) is over two orders of magnitudes larger than
that expected from star formation (Figure \ref{fig:lxlxsf}),
strongly suggesting an AGN component. The NE nucleus was
detected only in the soft X-rays. Taken at face value, its HR
and $\Gamma$ estimates suggest a steep spectrum; this, together
with its moderate hard X-ray luminosity ($L_{X,2-10\, {\rm
keV}}<3.0\times10^{41}$ erg s$^{-1}$) may indicate a source
dominated by star-formation-related processes. However, the
expected star-formation-related X-ray luminosities are lower
than the observed value ($14\pm10$ times fainter in the soft
band) or the upper limit (up to $\sim40$ times fainter in the
hard band), lending support for a low-luminosity AGN component
also in the NE nucleus. Similar to the case of the NW nucleus
of SDSS J1108+0659, the $U$-band image reveals intense star
formation activity in the NE nuclear region of SDSS J1146+5110.
While the apparent X-ray HR and $\Gamma$ estimates suggest no
significant absorbing column, the nuclear starburst component
may indicate otherwise.

\subsubsection{SDSS J1332+0606}

Both nuclei were optically classified as Type 2 Seyferts. The
NE nucleus was detected in both soft and hard X-rays, although
the counts were very low. The apparent HR and $\Gamma$
measurements may suggest a reflection dominated spectrum,
indicating the presence of a Compton-thick source. In the
Compton-thick scenario, the true intrinsic X-ray luminosities
would be much higher than our fiducial estimates, where only
modest absorption is assumed ($N_{\rm H} \sim
1.0^{+2.0}_{-0.99} \times 10^{22}$ cm$^{-2}$ estimated assuming
$\Gamma$=1.8, i.e., consistent with zero absorption). The
expected star-formation-related X-ray luminosities in both
bands are lower than the observed values within the
uncertainties, suggesting an AGN component. The SW nucleus was
undetected in the X-rays. While the upper limits of the X-ray
luminosities are consistent with an AGN component in the SW
nucleus as well, the possibility that only one AGN in the NE
nucleus ionizes gas in both galaxies cannot be ruled out.

In summary, our new \chandra\ and {\it HST} observations
support the binary-AGN scenario for two of our four targets
(SDSS J1108+0659 and SDSS J1146+5110). For the other two
targets (SDSS J1131$-$0204 and SDSS J1332+0606), the existing
data are still consistent with the binary-AGN scenario,
although the possibility of only one AGN ionizing both
components in the mergers cannot be fully ruled out.

\begin{figure*}
  \centering
    \includegraphics[width=190mm]{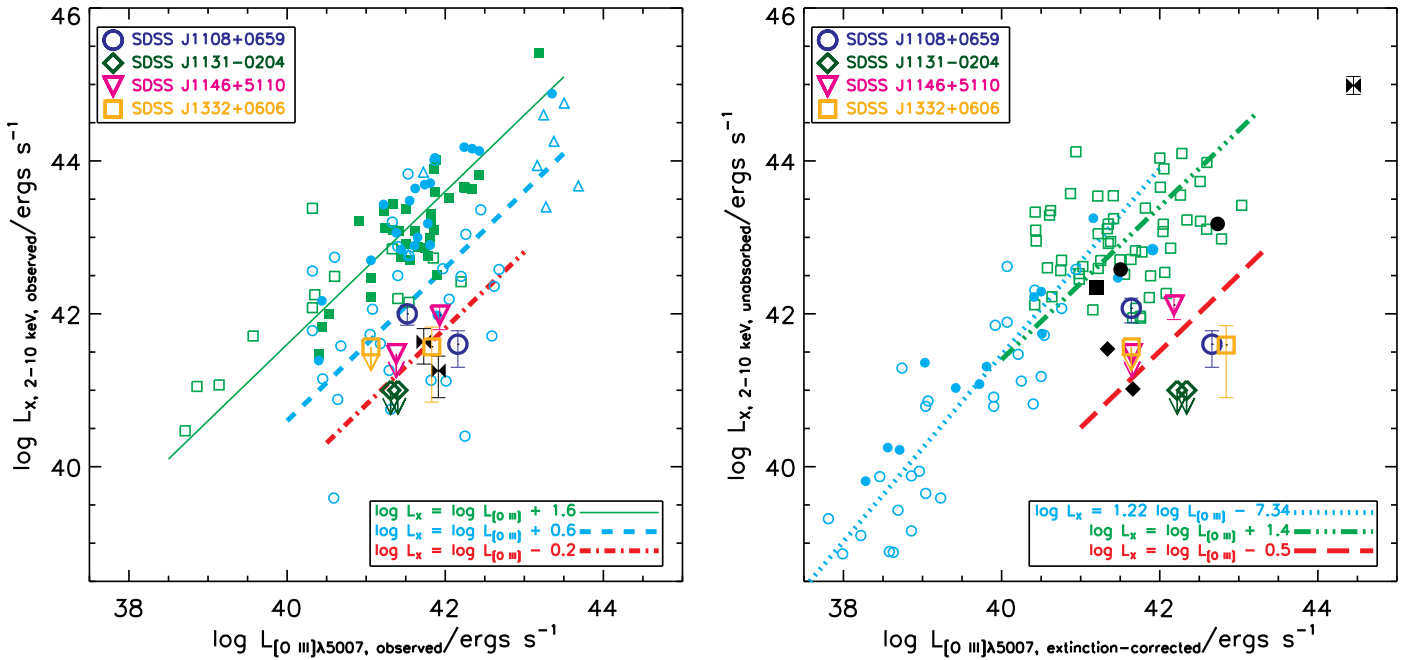}
    \caption{Hard X-ray vs. \OIIIb\ emission-line luminosity.
    Our targets are indicated with large colored open symbols, as denoted in the plot.
    Left panel: observed 2--10 keV
    luminosity vs. observed \OIII\ luminosity.
    The comparison samples include:
    hard X-ray selected AGNs
    (Type 1s as filled and Type 2s as open squares; both in green) and
    \OIII\ bright AGNs
    (Type 1s as filled and Type 2s as open circles; both in cyan) from \citet{heckman05},
    and optically selected Type 2 quasars (open upward triangles in cyan) from
    \citet{ptak06}.
    The red dash-dotted line is the mean relation for our targets,
    the cyan dashed line is for optically selected Type 2 AGNs \citep{heckman05},
    whereas the green solid line is for hard X-ray selected AGNs (both Type 1 and Type 2)
    and optically selected Type 1 AGNs \citep{heckman05}.
    Right panel: unabsorbed 2--10 keV luminosity vs. extinction-corrected \OIII\ luminosity.
    The comparison samples include:
    nearby optically selected Seyfert galaxies
    (Type 1 in filled and Type 2 in open circles; both in cyan) from \citet{panessa06},
    Type 2 Seyferts (green open squares) from the CSC-SDSS cross-match catalog of
    \citet{trichas12},
    and three previously known kpc-scale binary AGNs
    (NGC 6240 (a large filled square for the total nuclear emission; \citealt{ptak03})
    from \citealt{komossa03};
    Mrk 463 (large filled circles) from \citealt{mazzarella93} and \citealt{bianchi08};
    and Mrk 266 (large filled diamonds) from \citealt{brassington07}).
    The red long-dash line is the mean relation of our targets,
    the green dash-dot-dot line is for the CSC-SDSS sample,
    whereas the cyan dotted line is that from \citet{panessa06} for mixed Seyferts in
    nearby galaxies.
    Shown as filled bowties are two NLR-kinematics-candidate double-peaked \OIII\ AGNs
    (left panel: SDSS J1715+6008 from \citealt{comerford11a};
    and right panel: CXO J1426+35 (total emission) from \citealt{barrows12}).}
    \label{fig:lxlo3ratio}
\end{figure*}

\subsection{X-Ray-to-\OIII\ Luminosity Ratio}\label{subsec:x2oratio}

For optically selected Type 2 AGNs, the \OIII\ emission-line
luminosity is usually taken as a surrogate to estimate the
intrinsic hard X-ray luminosity, because the intrinsic AGN
continuum luminosity is obscured in the optical. Measurements
of the ratio $L_{X,{\rm 2-10\,keV}}/L_{{\rm [O\,{\tiny III}]}}$
for optically selected Type 2 AGNs span a wide range, with
values ranging from a few to a few hundred
\citep{mulchaey94,heckman05,panessa06}.

\subsubsection{Systematically Smaller X-Ray-to-\OIII\ luminosity Ratio
in Optically Selected kpc-scale Binaries than in Single AGNs}

In Figure \ref{fig:lxlo3ratio} we plot the hard X-ray
luminosity against the \OIII\ luminosity for each nucleus in
our targets. For context, we compare our targets to
observations of single AGNs and the few previously known
kpc-scale binary AGNs. We study both the relation between the
observed hard X-ray luminosity $L_{X,2-10\,{\rm keV,
observed}}$ and the observed \OIII\ luminosity $L_{{\rm
[O\,III],observed}}$, and that between the unabsorbed hard
X-ray luminosity $L_{X,2-10\,{\rm keV, unabsorbed}}$ and the
extinction-corrected \OIII\ luminosity $L_{{\rm
[O\,III],extinction-corrected}}$. We used different comparison
samples for the two as usually only the observed or the
corrected luminosity was available in any given sample.

For the $L_{X,2-10\,{\rm keV, observed}}$-$L_{{\rm
[O\,III],observed}}$ relation (left panel of Figure
\ref{fig:lxlo3ratio}), the comparison samples include the 47
hard X-ray (3--20 keV, in this particular case) selected AGNs
(the $z < 0.2$ subset of the \citealt {sazonov04} sample from
the {\it RXTE} all sky survey; \citealt{revnivtsev04}) and 55
optically selected local \OIII -bright AGNs
\citep{xu99,whittle92} studied by \citet{heckman05}, and 8
optically selected Type 2 quasars from \citet{ptak06}. The
\citet{heckman05} local AGN samples have similar redshifts and
$L_{{\rm [O\,III],observed}}$ to our targets, whereas the
\citet{ptak06} objects are at higher redshifts ($z\sim$0.3--0.8
compared to our targets at $z\sim$0.1--0.2) and have higher
$L_{{\rm [O\,III],observed}}$ (by $\sim1.5$ dex).
\citet{heckman05} showed that optically selected Type 2 AGNs
have systematically lower $L_{X,2-10\,{\rm keV, observed}}$ (by
an average of 1.0 dex) at a given $L_{{\rm [O\,III],observed}}$
than hard X-ray selected AGNs (both Type 1 and Type 2) and
optically selected Type 1 AGNs. Our optically selected
kpc-scale binary AGNs seem to have systematically smaller
$L_{X}/L_{{\rm [O\,III]}}$ (observed) values than even
optically selected single Type 2 AGNs, although the sample size
is still too small to draw a firm conclusion. The observed hard
X-ray luminosities of our targets on average are
$\sim0.8\pm0.2$ dex smaller at least\footnote{The upper limits
were included in the fit as detections.} than those of
optically selected single-nucleus Type 2 AGNs by
\citet{heckman05} at a fixed $L_{{\rm [O\,III],observed}}$
(i.e., average log($L_{X}/L_{{\rm [O\,III]}})$ = $-0.2$ with an
rms scatter of 0.40 dex, compared to log($L_{X}/L_{{\rm
[O\,III]}})$ = $0.6$ with an rms scatter of 1.1 dex for
optically selected single Type 2s).

For the $L_{X,2-10\,{\rm keV, unabsorbed}}$-$L_{{\rm
[O\,III],extinction-corrected}}$ relation (right panel of
Figure \ref{fig:lxlo3ratio}), the comparison samples include 47
Palomar Seyfert galaxies (optically selected Type 1 and Type 2
Seyferts drawn from the Palomar survey of nearby galaxies by
\citealt{ho95}) from \citet{panessa06}, and three previously
known kpc-scale binary AGNs (NGC 6240 from \citealt{komossa03}
and \citealt{ptak03}; Mrk 463 from \citealt{bianchi08}; and Mrk
266 from \citealt{brassington07}). We also include a new sample
of 55 single-nucleus Type 2 Seyferts from \citet{trichas12}.
This comparison sample was both optically and X-ray selected,
and was constructed by cross-matching the {\it Chandra} Source
Catalog \citep[CSC;][]{evans10} Release 1.1 with the SDSS DR7
spectroscopic galaxy catalog at $z<0.3$, and selecting objects
whose optical emission-line ratios \OIIIb /\hbeta\ and \NIIb
/\halpha\ are characteristic of Type 2 Seyferts according to
the \citet{kewley01} criteria.

\citet{panessa06} suggested that after properly accounting for
absorption correction (including for Compton thick sources),
optically selected Type 1 and Type 2 Seyferts all follow the
same $L_{X,2-10\,{\rm keV, unabsorbed}}$-$L_{{\rm
[O\,III],extinction-corrected}}$ relation. In particular,
optically selected Type 2 Seyferts, which were significantly
X-ray weaker than Type 1 Seyferts and quasars, also obey the
same relation, after the "Compton-thick" luminosity correction.
The average absorption correction for Compton-thick sources
(30\% of the sample) is $\sim$2--3 dex, or $\sim$0.5--1 dex for
the other Type 2 Seyferts.  Their absorption correction,
however, is significantly larger than that for our targets,
although our correction may have been underestimated given the
poor X-ray constraints.

After correction for gas absorption and dust extinction, the
unabsorbed hard X-ray luminosities of our targets appear to be
$\sim2.4\pm0.3$ dex smaller (at log$L_{{\rm [O\,III]}}$ of
42.0) than those expected from the \citet{panessa06} relation,
log$L_{X}$ = $1.22$log$L_{{\rm [O\,III]}}-7.34$, although the
absorption correction of our targets may have been
significantly underestimated (Section \ref{subsubsec:xspec}).
Our targets have average unabsorbed log($L_{X}/L_{{\rm
[O\,III]}})=-0.5$ (with an rms scatter of 0.69 dex), which is
$\sim1.9\pm0.3$ dex lower than that of the CSC-SDSS cross-match
comparison sample (average log($L_{X}/L_{{\rm [O\,III]}})=1.4$
with an rms scatter of 0.77 dex). Like our targets, the three
previously known kpc-scale binary AGNs (all of which are ULIRGs
and were discovered in the X-rays) also show smaller unabsorbed
log($L_{X}/L_{{\rm [O\,III]}})$ values than the
\citet{panessa06} relation.

\subsubsection{Interpretation: Higher Nuclear Gas Concentration
and/or Viewing Angle Effect}

We now discuss possible causes of the apparent hard X-ray weak
tendency in optically selected kpc-scale binary AGNs compared
to optically selected single Type 2 AGNs. Simulations suggest
that galaxy mergers may funnel significant amount of gas toward
galaxy centers, triggering both global and nuclear starburst
activity as well as AGN \citep[e.g.,][]{hernquist89,hopkins08}.
Using a sample of 1286 AGN pairs spectroscopically selected
from the SDSS with projected separations of a few kpc to a few
tens of kpc \citep{liu11a}, \citet{liu11b} have shown that the
fraction of both single and double AGNs stays constant with
decreasing projected separation at $>20$ kpc scales. This
suggests that at wide separations, the majority of the observed
AGNs are most likely due to stochastic accretion not associated
with tidal interactions \citep[e.g.,][]{ciotti07,ciotti10}.
However, the fraction of both single and binary AGNs increases
rapidly on scales below 20 kpc, indicating tidally enhanced AGN
in close galaxy pairs \citep[see also][]{ellison11}. In
addition, the fraction of binary AGNs in close galaxy pairs is
significantly higher than what would be expected from random
pairing of single AGNs. This, together with the correlation
observed in AGN luminosity between the merging components in a
pair \citep{liu11b}, provides direct evidence that the double
AGN observed in close galaxy pairs may indeed be connected to
merger activity. It is likely that the gas in kpc-scale binary
AGNs is being tidally funneled all the way to the nuclear
regions of both components, and either fuels the binary AGN
directly, or fuels a nuclear starburst which in turn feeds the
BHs.

We therefore suggest that a higher gas column in the nuclear
regions of kpc-scale binary AGNs, which is likely induced by
the merger events, may be at the root of the observed X-ray
weak tendency. Unlike X-ray emission which comes from the
accretion disk/corona of the accreting SMBHs, \OIII\ emission
in the NLRs comes from much larger scales
\citep{antonucci93,urry95}, and should therefore be much less
subject to nuclear obscuration. In this scenario, the X-ray
absorption column was significantly underestimated for our
targets, which is plausible given the large systematic
uncertainty of our X-ray measurement. While the three
previously known kpc-scale binary AGNs (which are all ULIRGs
and whose binary-AGN nature was all discovered in the X-rays)
also show smaller X-ray-to-\OIII\ luminosity ratios than single
Seyferts, the X-ray weak tendency seems to be less pronounced
than for our optically selected kpc-scale binary AGNs. This may
be due either to a selection effect, or a host-galaxy effect
(i.e., the nuclear gas distribution in ULIRGs -- disk-dominated
mergers -- is less concentrated than in optically selected
binary AGNs -- whose hosts have prominent stellar bulges -- due
to lower bulge-to-disk ratios of the host galaxies), or a
combination of both.

Alternatively, the X-ray weak tendency observed in our targets
may be caused by a viewing angle effect related to its
double-peak selection. The requirement of detecting
well-separated \OIII\ double peaks in velocity is likely to
pick out edge-on systems, and therefore would be biased toward
a higher absorbing gas column. For example, SDSS J1131$-$0204,
the only one of our targets which was undetected in the X-rays,
appears to show two stellar nuclei imbedded in a large edge-on
disk \citep{liu10b,shen10b}. To check whether the double-peak
selection causes the X-rays to be weak, we also examine two
double-peaked \OIII\ AGNs in the literature with available
\OIII\ and hard X-ray luminosity measurements (SDSS J1715+6008
from \citealt{liu10}, \citealt{smith09}, and
\citealt{comerford11a}, and CXO J1426+35 from
\citealt{barrows12}), whose emission-line profiles are more
likely caused by complex NLR kinematics, i.e., not caused by
kpc-scale binary AGNs. As shown in Figure \ref{fig:lxlo3ratio},
in both cases, the X-ray-to-\OIII\ ratios are smaller than the
\citet{heckman05} and \citet{panessa06} relations for the
observed and corrected luminosities, respectively. While this
comparison is based on only two objects, it may suggest that
the hard X-ray weak tendency is related to a double-peak
selection, regardless of whether there are two active BHs in
the system. However, the viewing angle effect is unlikely to be
the sole cause of the X-ray weak tendency, given that the three
previously known kpc-scale binary AGNs, which were not selected
from double-peaked narrow-line AGNs, also show smaller
X-ray-to-\OIII\ luminosity ratios than single Seyferts.

In summary, we conclude that the observed X-ray weak tendency
in our optically selected kpc-scale binary AGNs is likely
caused by a combination of (1) a higher nuclear gas column,
which may be induced by merger events, and (2) a viewing angle
bias related to its double-peak NEL selection. Of course,
another possibility is that we were simply unlucky with these
four systems, and they are not typical of kpc-scale binary
AGNs.

\section{Discussion}\label{sec:discuss}

We discuss the implications of our results for the general
double-peak narrow line selection approach of identifying
kpc-scale binary AGNs (Section \ref{subsec:imp_iden}), for the
attributes and limitations of optical identification compared
to X-ray searches (Section \ref{subsec:compare_withxray}), and
for the observed frequency of kpc-scale binary AGNs (Section
\ref{subsec:frequency}).

\subsection{On the Double-peak Approach for kpc-scale Binary-AGN
Identification}\label{subsec:imp_iden}

\subsubsection{X-Ray Confirmation of Optical Candidates: Success and Ambiguity}

Until recently, searches for kpc-scale binary AGNs have been
more or less serendipitous. Previous work
\citep{liu10b,mcgurk11,shen10b,fu11} has demonstrated that
selecting candidates based on double-peaked NELs and follow-up
with high spatial resolution imaging \citep{fu10,rosario11} and
spatially resolved spectroscopy \citep{comerford11b} is a
promising way to identify kpc-scale binary AGNs. With the new
\chandra\ and {\it HST} observations presented here, we have
critically examined the nature of the ionizing sources of the
optically identified kpc-scale binary AGNs. Our results confirm
the kpc-scale binary-AGN nature for two of the four optical
candidates; the data are still consistent with the binary
scenario for the other two, but we cannot rule out the
possibility of one AGN ionizing gas in both merging components.
While the result lends some further support to the overall
approach of systematically identifying kpc-scale binary AGNs
based on the double-peak selection, it also suggests that X-ray
confirmation of optical binary candidates can be challenging
and ambiguous, due to the complex nature of X-ray obscuration
in mergers.

\subsubsection{Importance of Identifying Double Nuclei Both in Gas and in Stars}

\citet[][see also \citealt{fu11}]{shen10b} suggested that the
majority of optically selected AGNs with double-peaked \OIIIc\
emission lines are not due to kpc-scale binary AGNs, but are
caused by complex NLR gas kinematics around single AGNs, such
as outflows \citep[e.g.,][]{rosario10} and/or rotating disks
\citep[e.g.,][]{smith12,blecha12}. \citet{shen10b} combined
high-resolution NIR imaging (to resolve the double stellar
bulges and to detect tidal features) with spatially resolved
optical spectroscopy (to locate the ionizing sources and to
register with the BHs/bulges) to identify strong kpc-scale
binary-AGN candidates \citep[see also][]{mcgurk11}. Detecting
two stellar components alone \citep[e.g.,][]{fu10,rosario11} is
not a sufficient condition for identifying binary AGNs, as the
double-peaked profile could be caused by NLR gas kinematics
around a single AGN in a merger. Neither is detecting spatial
offsets between the emission-line components alone
\citep[e.g.,][]{gerke07,comerford08,barrows12} a sufficient
condition, even though most double-peaked AGNs show spatial
offsets on kpc scales between the two velocity components
\citep{shen10b,fu11,comerford11b}, because such spatial offsets
are also expected and commonly observed in single AGNs due to
the spatial extent of the NLR, as demonstrated by
\citet{fischer11} in Mrk 78 \citep[see also][]{whittle04}.

We emphasize the importance of detecting double nuclei in both
gas and in stars in the identification of kpc-scale binary
AGNs. To pin down the ionizing sources, candidates from
spatially resolved optical spectroscopy can be confirmed using
imaging spectroscopy in the X-rays (as in this work) or in the
radio \citep{fu11b}. However, identifying the double stellar
bulges should be a prerequisite before carrying out X-ray or
radio observations. X-ray and/or radio observations would help
clarify the ambiguities for gas (i.e., in \OIII ), but cannot
substitute deep NIR imaging for stars. Double X-ray and/or
radio sources (as well as spatially resolved optical emission
regions coincident with double X-ray/radio sources) on a smooth
stellar background \citep[e.g., see the candidate reported
by][]{comerford11a}, is more likely to be complex NLR gas
kinematics \citep[e.g., a jet;][]{comerford11a} rather than
binary AGNs, as the case of Mrk 78. Exceptions to this included
binary AGNs with separations smaller than $\sim100$ pc, where
the two stellar bulges may have merged, and minor mergers,
which are difficult to resolve due to large contrast ratios
\citep[see the candidate example discovered by][]{fabbiano11}.

\subsubsection{Importance of High-resolution Deep Observations}

Our new \chandra\ and {\it HST} observations have unambiguously
confirmed the kpc-scale binary-AGN nature for SDSS J1108+0659
and SDSS J1146+5110. \citet{fu11} suggested that the case for
SDSS J1108+0659 was ambiguous from the detection of extended
emission-line regions (EELRs), using Keck NIR imaging assisted
with laser guide star AO and seeing-limited integral field
spectroscopy (IFS) on the University of Hawaii 2.2 m telescope.
However, the angular resolution of the IFS data (seeing
$\sim0.''8$) may have been insufficient to fully resolve the
close double nuclei in SDSS J1108+0659 (angular separation of
$0.''70$; Table \ref{table:obs}). More importantly, the
detection of EELRs on larger scales should not be taken as
evidence for or against the binary-AGN scenario, because EELRs
have been observed in both kpc-scale binary (such as in NGC
6240 and in Mrk 266) and single AGNs.

\citet{fu11} also suggested that SDSS J1146+5110 was a pair of
EELRs powered by a single AGN, but their IFS data were not
sensitive enough to detect the faint \OIII\ emission associated
with the NE nucleus, which was detected by \citet{liu10b} using
long-slit spectroscopy with the APO 3.5 m telescope.

\subsection{Comparing Optical to X-Ray Identification of kpc-scale Binary
AGNs}\label{subsec:compare_withxray}


We discuss our results in the context of comparing optical and
X-ray identification of kpc-scale binary AGNs. The
identification of AGNs based on any particular wavelength
window is likely to be limited by biases and incompleteness.
For example, \citet{heckman05} suggested that identifying AGN
based on bright \OIII\ emission lines will uncover the majority
of hard X-ray selected AGNs\footnote{Except for BL Lac objects
\citep{wolfe78} which contain very weak optical emission lines
\citep{collinge05}.}, whereas identifying AGN by hard X-rays
will miss a significant population of AGNs selected based on
the \OIII\ optical emission line. These hard X-ray faint yet
\OIII\ bright Type 2 AGNs are generally interpreted as heavily
obscured or even Compton-thick sources
\citep[e.g.,][]{risaliti99,bassani99,levenson02}.


Hard X-ray searches for kpc-scale binary AGNs are still limited
by low angular resolution and small number statistics. Low
resolution studies (e.g., by \citealt{jimenez07} using the
XMM-Newton \citep{jansen01}) are in general restricted to wide
(tens-of-kpc scales) pairs in the local universe, whereas high
resolution searches with \chandra\ are often confined to
relatively small samples \citep[e.g.,][]{teng05,teng12}. In
addition, X-ray searches often have to apply some pre-selection
based on optical AGN diagnostics to boost the success rate.

Similar to heavily obscured single AGNs, our result on the
hard-X-ray-to-\OIII\ luminosity ratio suggests that identifying
kpc-scale binary AGNs by hard X-rays is likely to miss a
population of optically selected Type 2 binary AGNs. To address
the fraction of such heavily X-ray-obscured binary AGNs among
all binary AGNs would require knowledge of the space densities
of Type-1-Type-1 and Type-1-Type-2 binaries, both optically
selected and hard X-ray selected, which is beyond the scope of
this paper. But given our result that the hard-X-ray-to-\OIIIb
-luminosity ratio of optically selected Type 2 binary AGNs
appears to be systematically smaller than single Type 2 AGNs,
we speculate that the obscured fraction could even be higher in
binary than in single AGNs.

On the other hand, the parent sample of double-peaked NEL AGNs
was selected to be Type 2 Seyferts \citep{liu10}. This, by
construction, will miss Type 1 binary AGNs\footnote{Our parent
AGN sample was selected from the SDSS DR7 main galaxy sample
\citep{strauss02}, supplemented with Type 2 quasars from
\citet{reyes08} that were not included in the main galaxy
sample. The main galaxy sample includes objects with redshifts
$z<0.7$ and spectral classification as galaxies by the
\textsf{specBS} pipeline \citep{SDSSDR6}, or quasars that were
targeted as galaxies. While \citet{hao05a} have shown that the
narrow emission line ratios of Type 1 AGNs also follow the
\citet{kewley01} criteria, our parent AGN sample will miss
(luminous) Type 1 AGNs which were not included in the DR7 main
galaxies.}, mixed Type 1/Type 2, as well as Type 2 binaries
involving LINERS and/or AGN-H {\tiny II} composites. In
particular, the two nuclei of the prototypical kpc-scale binary
AGN NGC 6240 \citep{komossa03} are both heavily obscured in the
optical, with emission-line ratios characteristic of LINERs
\citep{lutz99}; another example is Mrk 266, whose northern
nucleus is optically classified as a composite yet does contain
an X-ray AGN \citep{brassington07,mazzarella11}. \citet{liu11a}
have addressed the frequency of binary AGNs (including both
Type 1 and Type 2 objects as well as LINERs and composites) on
a few kpc to tens-of-kpc scales using spectroscopic galaxy
pairs from the SDSS DR7. Complementary to the double-peak
approach, the \citet{liu11a} sample includes both Type 1 and
Type 2 AGNs but is biased against binaries closer than 5 kpc in
projection. While more work is clearly needed to address the
frequency of kpc-scale Type 1 binaries and Type 2 binaries
containing LINERS/composites, the frequency of kpc-scale binary
AGNs inferred from the \citet{liu11a} sample is similar to that
from the double-peak approach \citep{liu10b,shen10b},
suggesting that the majority of kpc-scale binary AGNs are Type
2 objects, most of which are Seyferts, at least in the moderate
AGN-luminosity regime being considered.

\subsection{Frequency of kpc-scale Binary AGNs}\label{subsec:frequency}

Motivated by the serendipitous success in NGC 6240,
\citet{teng05} carried out a systematic search for binary AGNs
using \chandra\ imaging of a sample of eight (U)LIRGs which
contain double stellar nuclei; they found no additional strong
candidate. Similarly, \citet{teng12} carried out a \chandra\
surveys of a sample of 12 massive galaxy mergers, each of which
contains one optical AGN, but found no convincing case of
binary AGNs. The null results from such ``blind'' X-ray
searches in small samples are not unexpected, considering the
relatively low frequency of kpc-scale binary AGNs as measured
using optical identification based on much larger parent
samples \citep{liu10b,shen10b,fu11,liu11a}. The kpc-scale
binary AGN fraction from optical studies \citep[e.g.,
$\sim$0.5\%--2.5\%;][]{shen10b} is also consistent with the
X-ray estimate ($\sim$2\%) by \citet{koss12}, considering
statistical and systematic uncertainties.

Using follow-up NIR imaging and optical slit spectroscopy of 43
double-peaked \OIII\ Type 2 AGNs from the parent sample
presented in \citet{liu10}, \citet{shen10b} have estimated that
$\sim$0.5\%--2.5\% of Type 2 AGNs at $z<0.3$ are kpc-scale
binary AGNs of comparable luminosities, with a relative orbital
velocity $\gtrsim150$ km s$^{-1}$ \citep[see
also][]{liu10b,fu11,smith09,rosario11}. Using a complementary
approach based on a sample of 1286 AGN pairs with projected
separations $<100$ kpc and velocity offsets $<600$ km s$^{-1}$,
\citet{liu11a} have found a similar result of $\gtrsim$0.5\%
for the frequency of binary AGNs on kpc scales, although the
sample is biased against pairs with projected separations $<5$
kpc. These observed frequencies based on optical surveys have
been well reproduced by the phenomenological model presented by
\citet{yu11}. The model calculates the number density of binary
AGNs based on the observed galaxy merger rate and BH--bulge
scaling relations, under the assumption that significant
nuclear activity is triggered only in gas-rich mergers with
central massive BHs and only when the nuclei are roughly within
the half-light radius of each other.

The relatively low frequency of kpc-scale binary AGNs suggests
that optical identification combined with X-ray confirmation
may be more efficient than blind X-ray searches. While blind
hard X-ray surveys are needed to fully address the
incompleteness of optical identification
\citep[e.g.,][]{koss12}, a robust determination of the
frequency of kpc-scale binary AGNs using existing hard X-ray
surveys is still hampered by small sample statistics and poor
angular resolution. While searches in the radio using Expanded
Very Large Array and/or Very Long Baseline Array can probe much
smaller scales \citep[e.g.,][]{tingay11,lazio12}, the null
result from the comprehensive study of \citet{burke11}, based
on archival very long baseline interferometry observations of
3114 radio-luminous AGNs, may indicate that the detection yield
could be strongly limited by the requirement of both AGNs being
radio loud and the complication created by hot spots in radio
jets.

\section{Summary}\label{sec:sum}

We have presented {\it HST}/WFC3 F336W and F105W imaging and
\chandra\ ACIS-S 0.5--10 keV imaging spectroscopy of the four
optically selected kpc-scale binary AGNs identified by
\citet{liu10b}. We have further clarified the ambiguities
concerning the nature of the ionizing sources in optically
selected kpc-scale binary AGNs. We summarize our main findings
as follows.

\begin{enumerate}

\item[1.] By combining X-ray imaging spectroscopy and star
    formation constraints from high-resolution $U$-band
    imaging, we have critically examined the nature of the
    ionizing sources in the four optically selected
    kpc-scale binary AGNs. Our new \chandra\ and {\it HST}
    observations confirm the binary-AGN scenario for two of
    the four targets (SDSS J1108+0659 and SDSS J1146+5110).
    For the other two targets (SDSS J1131$-$0204 and SDSS
    J1332+0606), the existing data are still consistent
    with the binary-AGN scenario, although the possibility
    of only one AGN ionizing both components in the mergers
    cannot be ruled out. While the new observations lend
    some further support to identifying optical kpc-scale
    binary AGNs from a sample of optical candidates with
    double-peaked NELs, they also suggest that X-ray
    confirmation of optical binary candidates can be
    challenging and ambiguous, at least for heavily
    absorbed sources.

\item[2.] Combining our previous optical spectroscopy with
    the new X-ray observations, we have found tentative
    evidence for a systematically smaller
    hard-X-ray-to-\OIII\ luminosity ratio and/or higher
    Compton-thick fraction in optically selected kpc-scale
    binary AGNs than in optically selected single Type 2
    Seyferts. We suggest that the observed X-ray weak
    distinction may be caused by a combination of higher
    nuclear gas column (possibly induced by mergers) and
    viewing angle bias (related to the double-peak NEL
    selection).

\end{enumerate}

X-ray observations of more kpc-scale binary-AGN candidates are
clearly needed to put our results on a firmer statistical
ground. Building a much larger sample would also enable the
exploration of X-ray properties of kpc-scale binary AGNs as a
function of separation and host galaxy properties. While deeper
X-ray imaging of our targets would help better constrain their
X-ray spectral properties, the required observations for robust
spectral modeling (i.e., $>200$ counts) would be too expensive
(likely at least 10 times our exposure times for targets at
similar redshifts with comparable \OIII\ luminosities) to
justify even for a small sample. Using a different approach, we
are conducting a pilot imaging program with {\it Chandra} for a
few kpc-scale binary AGNs at lower redshifts drawn from the
parent sample of \citet{liu11a}. These observations will help
better address to what extent the X-ray weak tendency in
optically selected kpc-scale binary AGNs is caused by a viewing
angle bias related to double-peak narrow-line selection (i.e.,
as opposed to a merger-driven gas concentration effect).

The obscured nature of the four optically selected kpc-scale
binary AGNs allows us to study their host galaxy stellar
populations without much contamination from the AGN itself. In
paper II, we will use {\it HST} WFC3 F105W and F336W imaging to
characterize the detailed host galaxy morphologies and
small-scale star formation properties. We will also address the
properties and origins of the extended soft X-ray emission
detected in SDSS J1108+0659, to better understand starburst,
shock heating, outflows, and photoionization of gas on
galactic-wide scales in kpc-scale binary AGNs.

\acknowledgments

We thank Markos Trichas for his generous help with providing
the X-ray luminosity measurement of the CSC-SDSS cross-match
AGN catalog reported in \citet{trichas12}, and an anonymous
referee for a careful and useful report that improves the
paper.

Support for the work of X.L. was provided by NASA through
Einstein Postdoctoral Fellowship grant number PF0-110076
awarded by the {\it Chandra} X-ray Center, which is operated by
the Smithsonian Astrophysical Observatory for NASA under
contract NAS8-03060. Y.S. acknowledges support through the
Smithsonian Astrophysical Observatory from a Clay Postdoctoral
Fellowship. M.A.S. acknowledges the support of NSF grant
AST-0707266.

Support for this work was provided by NASA through Chandra
Award Number GO1-12127X issued by the \chandra\ X-ray Observatory
Center, which is operated by the Smithsonian Astrophysical
Observatory for and on behalf of NASA under contract NAS
8-03060. Support for program number GO 12363 was provided by
NASA through a grant from the Space Telescope Science
Institute, which is operated by the Association of Universities
for Research in Astronomy, Inc., under NASA contract NAS
5-26555.

This research has made use of software provided by the Chandra
X-ray Center in the application packages CIAO, ChIPS, and
Sherpa.

Funding for the SDSS and SDSS-II has been provided by the
Alfred P. Sloan Foundation, the Participating Institutions, the
National Science Foundation, the U.S. Department of Energy, the
National Aeronautics and Space Administration, the Japanese
Monbukagakusho, the Max Planck Society, and the Higher
Education Funding Council for England. The SDSS Web site is
http://www.sdss.org/.

The SDSS is managed by the Astrophysical Research Consortium for
the Participating Institutions. The Participating Institutions are
the American Museum of Natural History, Astrophysical Institute
Potsdam, University of Basel, University of Cambridge, Case
Western Reserve University, University of Chicago, Drexel
University, Fermilab, the Institute for Advanced Study, the Japan
Participation Group, Johns Hopkins University, the Joint Institute
for Nuclear Astrophysics, the Kavli Institute for Particle
Astrophysics and Cosmology, the Korean Scientist Group, the
Chinese Academy of Sciences (LAMOST), Los Alamos National
Laboratory, the Max-Planck-Institute for Astronomy (MPIA), the
Max-Planck-Institute for Astrophysics (MPA), New Mexico State
University, Ohio State University, University of Pittsburgh,
University of Portsmouth, Princeton University, the United States
Naval Observatory, and the University of Washington.

Facilities: {\it CXO} (ACIS), {\it HST} (WFC3), Sloan

\bibliography{binaryrefs}

\end{document}